\documentclass[%
 reprint,
runinaddress,
nofootinbib,
 amsmath,amssymb,
 aps,
pra,
]{revtex4-1}
\usepackage[dvipsnames]{xcolor}
\usepackage[english]{babel}
\usepackage{hyperref}
\usepackage{amsfonts}

\makeatletter
\def\bbl@set@language#1{%
  \edef\languagename{%
    \ifnum\escapechar=\expandafter`\string#1\@empty
    \else\string#1\@empty\fi}%
  \@ifundefined{babel@language@alias@\languagename}{}{%
    \edef\languagename{\@nameuse{babel@language@alias@\languagename}}%
  }%
  \select@language{\languagename}%
  \expandafter\ifx\csname date\languagename\endcsname\relax\else
    \if@filesw
      \protected@write\@auxout{}{\string\select@language{\languagename}}%
      \bbl@for\bbl@tempa\BabelContentsFiles{%
        \addtocontents{\bbl@tempa}{\xstring\select@language{\languagename}}}%
      \bbl@usehooks{write}{}%
    \fi
  \fi}
\newcommand{\DeclareLanguageAlias}[2]{%
  \global\@namedef{babel@language@alias@#1}{#2}%
}
\makeatother

\DeclareLanguageAlias{en}{english}

\usepackage{soul}
\usepackage{graphicx}
\usepackage{amsmath}
\usepackage{array}
\usepackage{amssymb}
\usepackage{dcolumn}
\usepackage{bm}

\usepackage{mathabx}
\usepackage{enumitem} 
\usepackage{tabularx}
\usepackage{soul}

\newcommand\blfootnote[1]{%
  \begingroup
  \renewcommand\thefootnote{}\footnote{#1}%
  \addtocounter{footnote}{-1}%
  \endgroup
}

\renewcommand{\vec}[1]{\mathbf{#1}}

\def\vec#1{\boldsymbol{#1}}

\def\pd2v#1#2#3{\frac{\partial^2 #1}{\partial #2 \partial #3}}

\def \vec#1{\mathbf{#1}}
\def \2x2mat#1#2#3#4{
\left( \begin{array}{cc}
#1 &  #2 \\  #3 &  #4
\end{array} \right)
}

\begin{document}

\preprint{APS/123-QED}

\title{Quantum Light Detection and Ranging}

\author{Jiuxuan Zhao$^{\,1}$}%

\author{Ashley Lyons$^{\,2}$}%

\author{Arin Can Ulku$^{\,1}$}

\author{Hugo Defienne$^{\,2}$}

\author{Daniele Faccio$^{\,2,*}$}%

\author{Edoardo Charbon$^{\,1,*}$}%
\affiliation{ \\ $^{1}$Advanced Quantum Architecture Laboratory (AQUA), Ecole Polytechnique Federale de Lausanne (EPFL), 2002 Neuchatel, Switzerland \\ $^{2}$School of Physics and Astronomy, University of Glasgow, Glasgow G12 8QQ, UK \
}%

\date{\today}
\begin{abstract}
Single-photon light detection and ranging (LiDAR) is a key technology for depth imaging through complex environments. Despite recent advances, an open challenge is the ability to isolate the LiDAR signal from other spurious sources including background light and jamming signals. Here we show that a time-resolved coincidence scheme can address these challenges by exploiting spatio-temporal correlations between entangled photon pairs. We demonstrate that a photon-pair-based LiDAR can distill  desired  depth information in the presence of both synchronous and asynchronous spurious signals without prior knowledge of the scene and the target object. This result enables the development of robust and secure quantum LiDAR systems and paves the way to time-resolved quantum imaging applications.  
\end{abstract}

\maketitle

\blfootnote{$^*$Corresponding authors:\\
  daniele.faccio@glasgow.ac.uk\\
  edoardo.charbon@epfl.ch \\ }

Light detection and ranging (LiDAR) systems with the ability to reach long distance at high speed and accuracy have emerged as a key technology in autonomous driving, robotics, and remote sensing~\cite{schwarz2010mapping}. Today miniaturised LiDARs are integrated in many consumer electronics devices, e.g. smartphones. Moving beyond depth sensing, the LiDAR technique has been also used for non-line-of-sight imaging~\cite{velten2012recovering,gariepy2016detection,o2018confocal,liu2019non,faccio2020non}, imaging through scattering media~\cite{lyons2019computational} and biophotonics applications~\cite{bruschini2019single}. A typical LiDAR system records the time-of-flight, $t$, of  light back-reflected from a scene, which enables to estimate distance $d=ct/2$, where  $c$ is the speed of light~\cite{sun2016single}.  Thanks to their single-photon sensitivity, picosecond temporal resolution and low cost, single-photon avalanche diodes (SPADs) are widely used as detectors in LiDAR~\cite{shin2016photon,tachella2019real}. In this respect, two well established techniques can be used to achieve the timing information at picosecond resolution:  time-correlated single-photon counting (TCSPC) that operates by recording a time-stamp for each individual photon~\cite{zhang201830,ximenes_JSSC_2019,choi_JSSC_2021} or  time gating  in which a gate window is finely shifted~\cite{ulku2018512,morimoto2020megapixel,ren2018high,chan2019long}.\\
Despite recent advances, interference is a major challenge for robust and secure LiDAR applications through complex environments. In our work, the term `interference' refers to the detection by the LiDAR sensor of any optical signals other than those emitted by the LiDAR source. These may originate from ambient light, other LiDAR systems operating concurrently and intentional spoofing signals. In addition to depth distortion such as degradation in accuracy and precision,  interference could result in misleading information, causing the system to make incorrect decisions. Over the past several years, some approaches addressing LiDAR interference have been proposed. One technique isolates the signal based on temporal correlations between two or more photons~\cite{niclass_JSSC_2013}, which effectively suppresses noise due to ambient light. Another technique based on laser phase modulation can reduce both ambient light and mutual interference~\cite{ximenes_JSSC_2019,choi_JSSC_2021}. However, these approaches have limitations. For example, an external signal can still spoof the LiDAR detector if it copies the temporal correlation or phase modulation of the illumination source, which is easily achievable by placing a photodiode close to the target object. To date, there is no LiDAR system  immune to all types of interference.\\
The use of non-classical optical states can also improve object detection in the presence of spurious light and noise. In a quantum illumination protocol, a single photon is sent out towards a target object while its entangled pair is retained and used as an ancilla~\cite{lloyd2008enhanced}. Coincidence detection between the returned photon and its twin increases the effective signal-to-noise ratio compared to classical illumination, an advantage persisting even in the presence of noise and losses. Recently, practical quantum illumination schemes have been experimentally demonstrated using spatially entangled photon pairs for target detection~\cite{lopaeva2013experimental,zhang2020multidimensional} and imaging objects~\cite{england2019quantum,defienne2019quantum,gregory2020imaging} in the presence of background light and spurious images. These approaches rely on the ability to measure photon coincidences between many spatial positions, which is conventionally performed using single-photon sensitive cameras such as electron multiplied charge coupled device (EMCCD) cameras~\cite{moreau2012realization,edgar2012imaging,defienne2018general}, intensified(i)CCD~\cite{chrapkiewicz2014high,chrapkiewicz2016hologram} and SPAD cameras~\cite{eckmann2020characterization,ianzano2020fast,ndagano2020imaging,defienne2021full}. However, while a handful of works have reported the use of photon pairs for  target detection at distance~\cite{liu2019enhancing,frick2020quantum,ren2020time}, no imaging LiDAR experiments with absolute range distance and interference or background rejection  have been reported.\\
\begin{figure*}
\includegraphics[width=1 \textwidth]{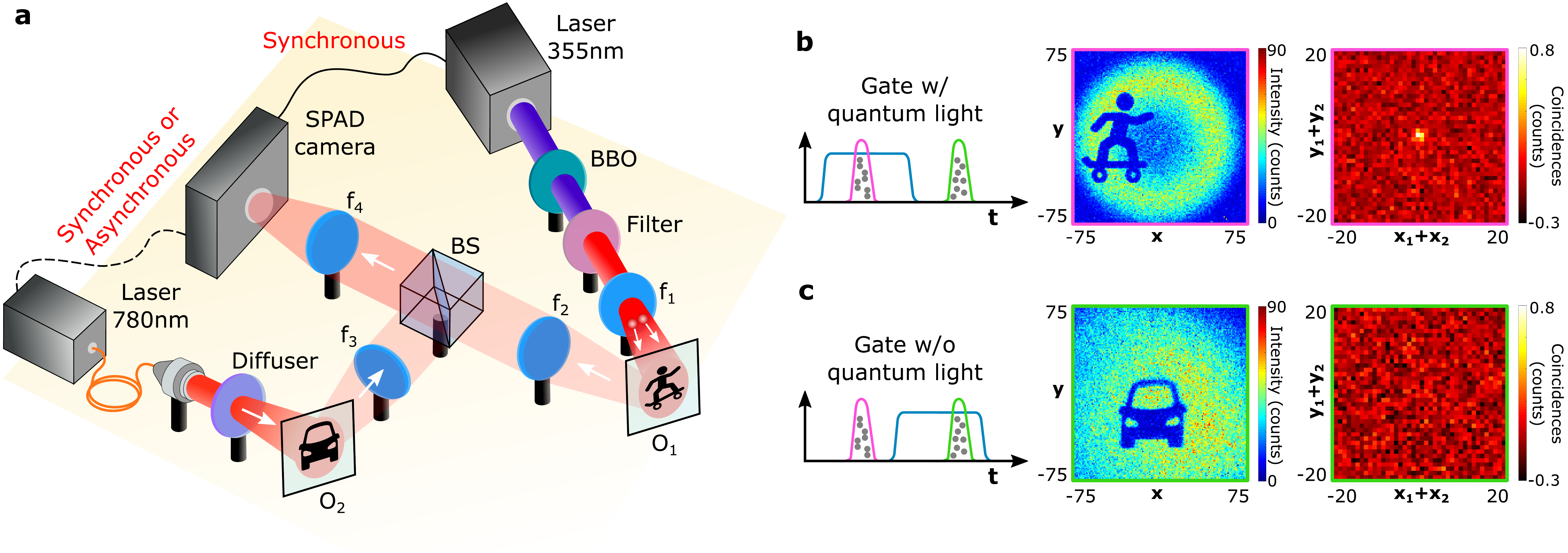} \caption{ \textbf{Experimental setup and principle.} \textbf{(a)} An object $O_1$ placed in the far field of a 1-mm-thick $\beta$-Barium Borate (BBO) nonlinear crystal is illuminated by spatially entangled photon pairs produced via type-I spontaneous parametric down conversion (SPDC), while an object $O_2$ is illuminated by diffused classical light. A lens $f_1 = 50$ mm is positioned after the crystal to direct the photon-pairs towards $O_1$. Both objects are composed of an absorptive pattern layer on a reflective surface. They are imaged onto the SPAD camera using lens $f_2 = 100$ mm, $f_3 = 50$ mm, $f_4 = 100$ mm and an unbalanced beam splitter (0.1R/0.9T).  \textbf{(b)} When the SPAD gate window is set to capture only photon pair pulses reflected by the quantum object, the ``skater-shape'' object appears in the intensity image and a peak is detected in the spatially-averaged correlation image, which shows the number of photon coincidences spatially averaged over all pair of pixels $\vec{r_1}$ and $\vec{r_2}$ separated by a given distance $\vec{r_1}+\vec{r_2}$. The correlation peak confirms the presence of photon pairs among the detected photons. \textbf{(c)} When the SPAD gate window is set to capture only classical light, the ``car-shape'' object appears in the intensity image whereas no peak obtained is visible in the spatially-averaged correlation image. Intensity and spatially-averaged correlation images were reconstructed from $N = 2000$ frames (8-bit) acquired using an exposure time of $350$ ns (1-bit). Intensity image coordinate units are in pixels. 
\label{Figure1}}
\end{figure*}
In this work, we demonstrate a {quantum}   LiDAR system immune to any type of classical interference by using a pulsed light source of spatially entangled photon pairs and a time-resolved SPAD camera. We use spatial anti-correlations between photon pairs as a unique identifier to distinguish them from any other light sources in the target scene. In particular, we show our LiDAR system successfully images objects and retrieves their depths in two different interference scenarios mimicking the presence of spoofing or additional classical LiDAR signals. In the first case, spurious light from a synchronised laser is used to demonstrate the robustness against intentional spoofing attacks. In the second case, the interference takes the form of asynchronous pulses imitating the presence of multiple background LiDAR systems running in parallel. The results show that our approach enables to image with high depth resolution while offering immunity to classical light interference.\\
\textbf{Imaging system.} The experimental setup is shown in Fig.~\ref{Figure1}a. Spatially entangled photon pairs are produced by type-I spontaneous parametric down conversion (SPDC) with a $\beta$-Barium Borate (BBO) nonlinear crystal pumped by a $355$ nm pulsed laser with a repetition frequency of $20$ MHz. {The objects to be imaged are masks placed on a reflective mirror}. One object $O_1$ (“skater”) is placed in the far field of the crystal thus the down-converted photon pairs are spatially anti-correlated at the object plane. Another object $O_2$ (“car”) is illuminated by a diffused $780$ nm laser pulsed at $20$ MHz as well to produce the interference. Both objects are imaged onto the SPAD camera \textit{SwissSPAD2}~\cite{ulku2018512} (see Methods). Similar to a typical LiDAR scheme, the pump laser is synchronised with the camera, while the spoofing signal generated by a classical pulse laser can be synchronous or asynchronous.\\ 
\begin{figure*}
\includegraphics[width=1 \textwidth]{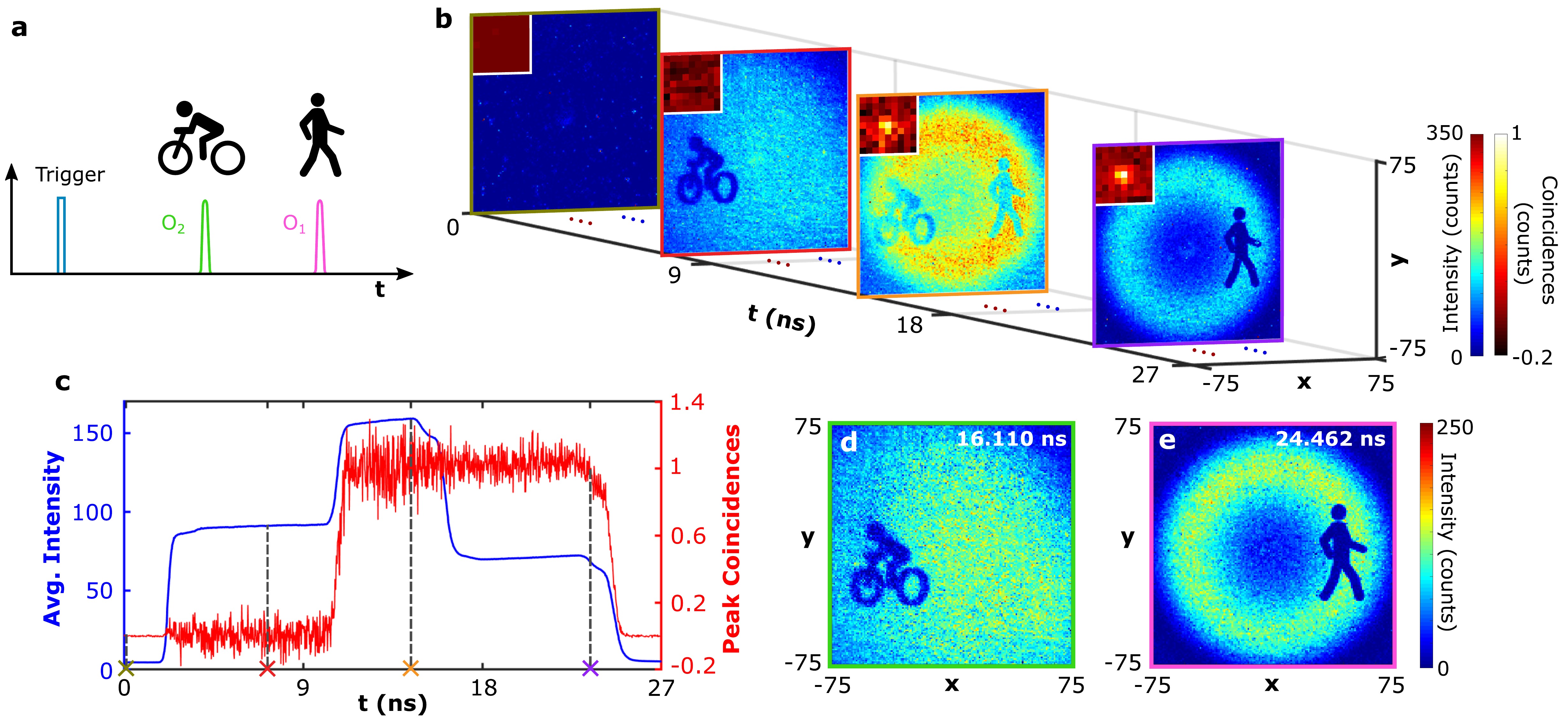} 
\caption{\label{Figure2} \textbf{Results with synchronous classical light interference.}    \textbf{(a)} The reflected light from objects $O_1$ (``person") and $O_2$ (``bike") are both synchronous with the camera. \textbf{(b)} shows the selected intensity and spatially-averaged correlation images ($9\times9$ central data) at the gate positions $0.09$ ns, $7.2$ ns, $14.58$ ns and $23.4$ ns covered with none reflected light, reflected light from only $O_1$, $O_1$ $\&$ $O_2$, and only $O_2$ respectively. Correlation peaks are obtained at $14.58$ ns and $23.4$ ns gate positions when there is quantum light reflected to the camera. The measurement is implemented over a time range of $27$ ns corresponding to $1500$ continuous gate positions with a proper time offset initially to the pump laser trigger. \textbf{(c)} Average intensity over all pixels (blue curve)  and the peak coincidences (red curve) values along the measured time range. The four positions in \textbf{(b)} are also marked on the horizontal axis of the curve. \textbf{(d)} The subtracted intensity image of $O_2$ (classical) and its arrival time ($16.110$ ns) to the camera by locating the first falling edge of the average intensity profile. \textbf{(e)} The subtracted intensity image of $O_1$ (quantum) and its arrival time ($24.462$ ns) to the camera by locating the falling edge of the correlation peak profile. Experiments are performed by $N = 5000$ frames (8-bit) acquired in $13.5$ s at each gate position using an exposure time of $350$ ns for 1-bit frame. The time step between two successive gate windows is $18$ ps. Intensity image coordinate unites are in pixels.}
\end{figure*}
\begin{figure*}
\includegraphics[width=0.9 \textwidth]{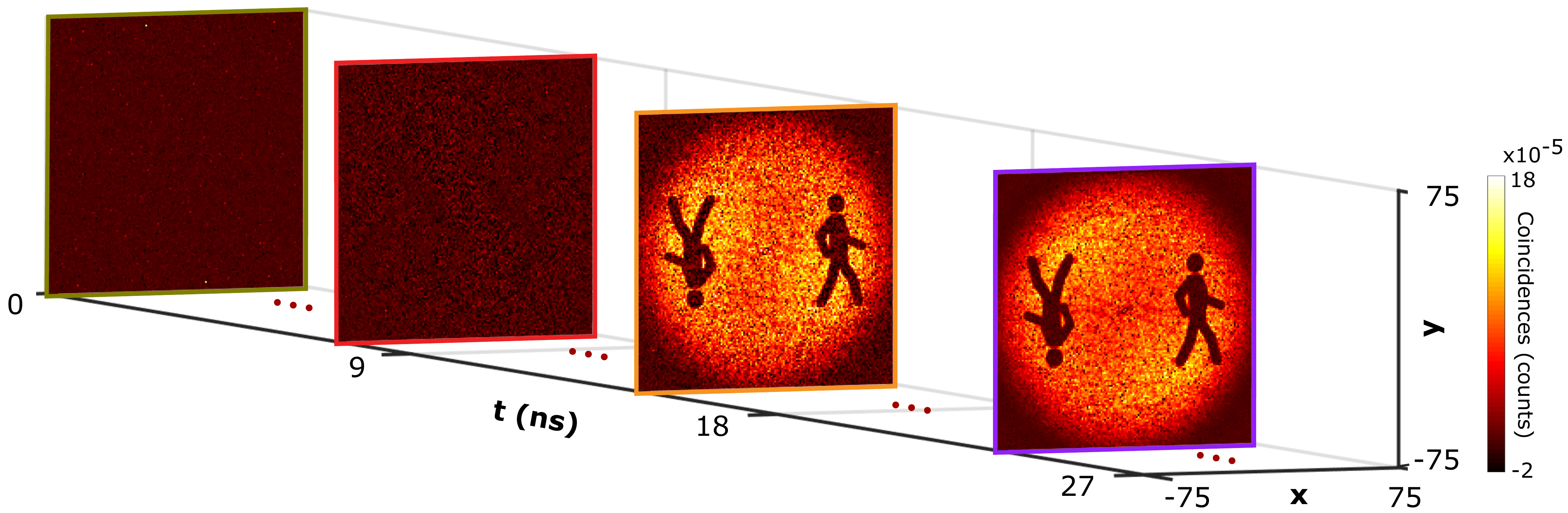} 
\caption{\label{Figure3} \textbf{Measurement of spatially-resolved correlation images over time.} The shape of the object illuminated by photon pairs (``person") is retrieved by measuring spatially-resolved correlation images at the gate positions $0.09$ ns, $7.2$ ns, $14.58$ ns and $23.4$ ns. Each image is obtained by acquiring $5$ million frames (8-bit), which corresponds to approximately $3.8$ hours of acquisition. See methods for more details.}
\end{figure*}
As in conventional time-gated LiDAR, backscattered photons with specific time-of-flight are detected by scanning a gate window ($15$ ns wide) using $18$ ps time steps, which corresponds to a depth resolution of $2.7$  mm~\cite{ulku2018512,morimoto2020megapixel}. At each gate position, a set of 8-bit frames ($\sim10^3$ frames) is acquired to reconstruct two different types of images: (i) a classical intensity image, obtained by summing all frames, and (ii) a spatially-averaged  photon  correlation image computed by identifying photon coincidences in the frame set using a technique detailed in~\cite{defienne2018general} (see Methods). The intensity image retrieves the shape of the objects in the scene at a given depth, while the spatially-averaged correlation image measures spatial correlations between detected photons  to identify the presence of photon pairs. For example, if only reflected photon pairs are captured within the gate window (Fig.~\ref{Figure1}b), the intensity image shows the “skater” object and an intense peak is observed at the center of the spatially-averaged correlation image. The presence of such a correlation peak above the noise level confirms the presence of photon pairs among the detected photons. If only classical light is detected (Fig.~\ref{Figure1}c), the intensity images show the “car” object illuminated by the pulse laser and the spatially-averaged correlation image is flat. \\
\begin{figure*}
\includegraphics[width=1 \textwidth]{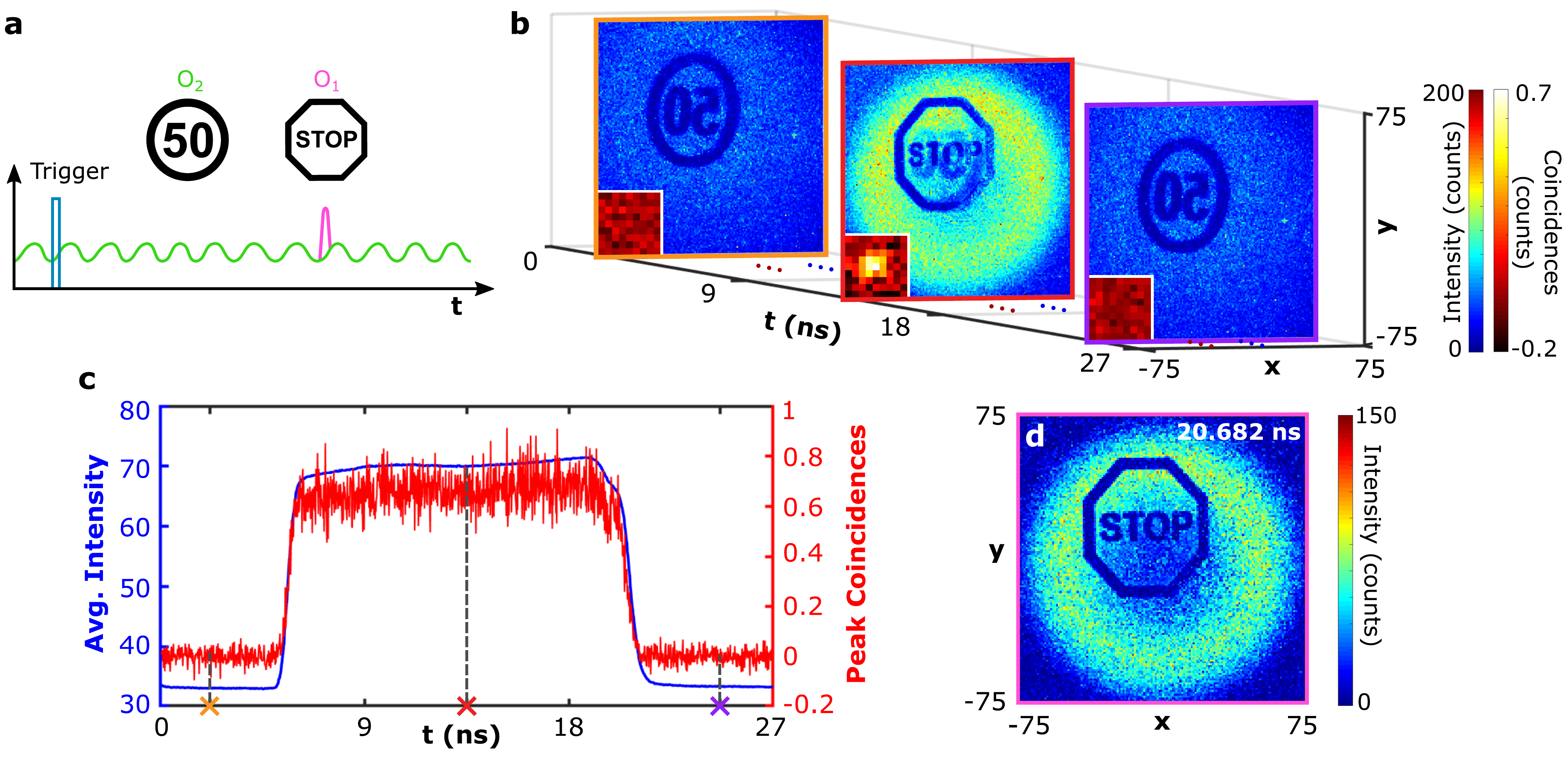} 
\caption{\label{Figure4} \textbf{Results with asynchronous spurious light.} \textbf{(a)} Photon pairs reflected  by object $O_1$ (``STOP traffic sign") is synchronous with the camera, while classical photons reflected by $O_2$ (``50 traffic sign") arrives at the camera in temporally random sequences as the classical laser is asynchronous. \textbf{(b)} The camera scanned over a time range of $27$ ns (1500 continuous gate positions). Intensity and spatially-averaged correlation images (central 9$\times$9 pixels area) are shown for three different gate positions ($2.16$ ns, $13.5$ ns and $24.66$ ns). The correlation peak only appears at the gate window covered with photon pairs reflected by $O_1$. \textbf{(c)} The corresponding three gate positions are marked in the curve of the average intensity and correlation peak responses over the detected time range. \textbf{(d)} Intensity image reconstructed by subtracting intensity image at $13.5$ ns by this at $24.66$ ns. At each gate position, $N = 3000$ frames (8-bit) were acquired in $8.1$ s using an exposure time of $350$ ns (1-bit). The time step between two successive gate positions is $18$ ps. Intensity image coordinate unites are in pixels.}
\end{figure*}
\textbf{Synchronous classical light interference}. First, we consider the case of a spurious classical source of light that is synchronised with the SPAD camera i.e. photons reflected by both $O_1$ (``person") and $O_2$ (``bike") are synchronous with the camera (Fig.~\ref{Figure2}a). This scenario corresponds to a spoofing attack. To operate the LiDAR, the gate window is continuously shifted over a range of $27$ ns, which corresponds to 1500 gate positions. Figure~\ref{Figure2}b shows the  intensity and spatially-averaged correlation images (zoom $9\times9$ pixels in inset) measured at four specific gate positions $0.09$ ns, $7.2$ ns, $14.58$ ns and $23.4$ ns. At the early gate position ($0.09$ ns), there is only noise recorded by the camera such as dark count, crosstalk, afterpulsing and ambient light. As the gate window is shifted, $O_2$ appears in the intensity image ($7.2$ ns) and the absence of a peak in the spatially-averaged correlation image shows that it originates from classical light alone. When the reflected quantum light starts to be collected in the gate window together with the classical light ($14.58$ ns), $O_1$ and $O_2$ are superposed in the intensity image and a correlation peak becomes visible. For the late gate window ($23.4$ ns) the classical laser pulse vanishes while only quantum light is detected, as shown by the peak persisting in the spatially-averaged correlation image, and only $O_1$ is visible in the intensity image.\\
To acquire depth information and distinguish classical interference, the spatially-averaged intensity and correlation peak values represented in function of the gate position in Fig.~\ref{Figure2}c are analyzed. The two-step average intensity profile represents the double reflections from $O_1$ and $O_2$, while the correlation peak profile only reveals the trend of quantum light over the given time range. By locating the falling edges of the intensity profile, the arrival time information of all the objects can be obtained~\cite{morimoto2021superluminal,chan2019long}. Whilst by just searching for the last falling edge of the correlation peak profile, the arrival time information of the quantum object is extracted. As shown in Fig.~\ref{Figure2}d the arrival time of the classical object of $16.110$ ns and its intensity image are obtained. The arrival time of the quantum object, $24.462$ ns, is located at the last fitted falling edge of the correlation peak profile and the corresponding intensity image is subtracted in Fig.~\ref{Figure2}e. Refer to the Supplementary Video for the scanned results over the entire detected range. The proposed dual-profile locating method enables locating and distinguishing objects illuminated by quantum light or classical light. \\
{The anti-spoofing capability works as described, as long as the time delay between the two objects is larger than the temporal resolution of the SPAD camera. One may then enhance the removal of temporally overlapping interferences by increasing the number of frames (e.g. up to $\sim 10^6$ 8-bit frames) for each gate delay} so as   to retrieve a spatially-resolved correlation image instead of a spatially-averaged correlation image~\cite{defienne2018general,defienne2019quantum,gregory2020imaging,defienne2021full}. An example is shown in Fig.~\ref{Figure3}: a spatially-resolved correlation image retrieves directly the shape of the object illuminated by photons pairs and remains insensitive to classical interference (classical background noise added in the experiment).  In fact, such a spatially-resolved correlation image could potentially be measured at all gate positions of the LiDAR scanning. However, the acquisition time is much longer than that required to retrieve an spatially-averaged correlation image (a few hours instead of seconds for a single time gate delay) and it is therefore better to limit its use to gate positions for which the objects cannot be distinguished otherwise. In addition, note that because of the anti-symmetric spatial structure of photon pairs illuminating the object, each spatially-resolved correlation image contains both the object and its symmetric image, which means that the object must interact with only half of the photon pair beam to be imaged through correlations without ambiguity.\\
\textbf{Asynchronous classical light interference.} In real-world applications, another possible scenario is the interference coming from ambient light and other LiDAR systems. We therefore consider a classical source of light that is not synchronised with the SPAD camera but still running at the same repetition frequency ($20$ MHz) and illuminates the object $O_2$ “50 traffic sign" (Fig.~\ref{Figure4}a). Fig.~\ref{Figure4}b shows the intensity and correlation  images at three example gate positions $2.16$ ns, $13.5$ ns and $24.66$ ns. The object $O_2$ is visible in the intensity images at all gate positions as background noise. When the gate is shifted to $13.5$ ns, the SPAD also captures photon pairs reflected by $O_1$ (“STOP traffic sign") and both objects are superimposed in the intensity image. We now also observe a peak in the spatially-averaged correlation image which highlights the presence of photon pairs. By locating the falling edge of the spatially-averaged correlation peak shown in Fig.~\ref{Figure4}c, the time arrival information of the quantum object is located at $20.682$ ns, and its intensity image is also obtained by subtraction as shown in Fig.~\ref{Figure4}d. 
See the Supplementary Video for the entire measured results. \\
\textbf{Discussion.} We demonstrated a LiDAR system based on spatially entangled photon pairs showing robustness against interference from classical sources of light. In particular, we showed its successful use in the presence of (i) a spoofing attack (synchronous classical light interference) and (ii)  of a background light and another LiDAR system operating in parallel (asynchronous classical light interference). Note also that because the quantum LiDAR harnesses anti-correlations between photon pairs to retrieve images, it is also immune to classically-correlated sources of light such as thermal and pseudo-thermal sources in which photons are position-correlated~\cite{valencia_twophoton_2005}.  \\
In our current implementation, time gate position is acquired in several seconds so that the full scanning takes several hours (5.6 hours for the synchronous case and 3.4 hours for the asynchronous case). This total acquisition time can however be significantly decreased by reducing (i) the acquisition time per gate position  and (ii) the number of gate positions to detect the falling edge of quantum light (currently $1500$ steps scanned linearly). For example, in the case of the synchronous classical light shown in Fig.~\ref{Figure2}, the quantum illuminated object could be located by measuring only $300$ frames for $8$ different gate positions by using a correlation-driven scanning and falling edge fitting algorithm, which would reduce the total acquisition time to $7$ seconds (see details in the Supplementary Information). In addition, the speed of the SPAD camera in our experiment was limited to $370$ fps by the readout architecture, but it has been demonstrated that the similar cameras can be operated at frame rates up to $800 000$ fps~\cite{ISSCC2018}, which would further diminish the total acquisition time and potentially reach real-time acquisition. Furthermore, in the current quantum LiDAR prototype the target object is a two dimensional `co-operative' object attached to a mirror, which ensures enough photon pairs are reflected and collected by the camera. However, the proposed scheme can be extended to scattering materials with three dimensional profiles by using brighter photon pair sources and more sensitive SPAD cameras, which are currently under development. Looking forward, these results could enable the development of robust and secure LiDAR systems and more general time-resolved quantum imaging applications.

\section*{Methods} 

\noindent \textbf{Details of the experimental setup.} The $355$ nm pump laser used in this work is VisUV-355-HP (Picoquant GmbH) with $27$ mW average output power at a repetition frequency of $20$ MHz. The non-linear crystal cut for type-I SPDC is a $\beta$-Barium Borate crystal of size  $5\times5\times1$ mm with a half opening angle of 3 degrees (Newlight Photonics). A $710$ nm bandpass filter is placed after the BBO crystal to filter out the spurious pump light. The $780$ nm classical laser (PiL079XSM, ALS GmbH) running at a repetition rate of $20$ MHz with $40$ ps (FWHM) pulse width is coupled with a fiber and then connected to a collimator, the output power of which can be finely tuned. A diffuser (ED1-C50-MD, Thorlabs) is used to create even illumination of the classical light over the object. The $355$ nm pump laser always operates as a master to trigger only the camera in the asynchronous measurement, and both camera and the $780$ nm laser in the synchronous experiment.\\
\noindent \textbf{Details of the SPAD camera.} The camera used in our study is the \textit{SwissSPAD2} with microlens on chip. It is composed of $512\times512$ pixels with a pitch of $16.38$ $\mu$m, a native fill factor of $10.5\%$ and photon detection probability (PDP) of approximately $25\%$ at $700$ nm. The camera runs in time-gating mode by scanning gate windows ($\sim15$ ns wide) continuously with a time step of $18$ ps. The starting gate position is tuned to be prior to the first synchronous light reflection by adding an appropriate initial offset to the laser trigger. There is no initial calibration approach implemented in this work to compensate the time-arrival skew from electrical signal (cable length, trigger circuit) and optical signal (fiber) as we only focus on the relative range of different objects. At each gate window, the number of N $8$-bit frames are transferred to the computer by USB$3$ connection and photon correlations are processed on a GPU before the gate shifts to the next successive position to run the same operation. Each 8-bit frame is accumulated by 255 successive 1-bit measurements with $350$ ns exposure time. The overall acquisition speed is $370$ fps (with $10.2$ $\mu$s readout time for each bit) and {the post-processing time is less than $1$ ms per 8-bit frame when running on a GPU}. The central $150\times150$ pixels are selected for all experiments to minimize the skew influence due to the electrical propagation across the SPAD array. The gate control signal injected to the middle of the pixel array resulted a symmetrical time propagation to the right and left pixels. To remove the hot pixels, we define a threshold at $200$ dark counts and set all pixel values above this threshold in each frame to 0, as described in~\cite{defienne2021full}. Then the hot pixels are smoothed by its neighboring pixels~\cite{shin2016photon} for all intensity images shown in the work. The crosstalk effects are also removed by setting the correlation values from direct neighbour pixels to 0. \\
\noindent \textbf{Spatial correlation image calculation.} The photon coincidence processing model used in this study is detailed in ~\cite{defienne2018general}, in which the spatial joint probability distribution (JPD) $\Gamma$($\vec{r_i}$,$\vec{r_j}$) of entangled photon pairs is measured by multiplying values measured at pixel $i$ in each frame by the difference of values measured at pixel $j$ between two successive frames:
\begin{equation}
\label{model1}
\Gamma(\vec{r_i},\vec{r_j}) = \frac{1}{N} \sum_{\ell=1}^N [ I_\ell(\vec{r_i}) I_\ell(\vec{r_j}) -I_\ell(\vec{r_i}) I_{\ell-1}(\vec{r_j})],
\end{equation}
where $N$ is the acquired number of frames at each gate position.  $I_\ell(\vec{r_i})$ and $I_\ell(\vec{r_j})$ represent the photon-count value at any pixel $i$ and $j$ (in position $\vec{r_i}$ and $\vec{r_j}$) of the $\ell^{th}$ frame ($\ell \in [\![1;N]\!]$). The genuine coincidences that only originates from correlations between entangled photon pairs are obtained by removing the accidental coincidences resulting from dark counts, after-pulsing, hot pixels, crosstalk, detection of multiple photon pairs and stray light. Both the (i) spatially-averaged correlation and the (ii) spatially-resolved correlation image used in our study can be extracted from the JPD: \\
(i) The spatially-averaged correlation image (noted $\Gamma_+$) is calculated from JPD $\Gamma$ using the formula:
\begin{equation}
\Gamma_+(\vec{r_1}+\vec{r_2}) = \sum_{\boldsymbol{\vec{r}}} \Gamma(\vec{r_1}+\vec{r_2}-\vec{r},\vec{r}).
\end{equation}
This represents the average number of photon coincidences detected between all pairs of pixels $\vec{r_1}$ and $\vec{r_2}$ separated by a distance $\vec{r_1}+\vec{r_2}$.\\
(ii) The spatially-resolved correlation image is defined as the anti-diagonal component of the JPD $\Gamma(\vec{r},-\vec{r})$. It represents the number of photon coincidences detected between symmetric pair of pixels. 

\bibliographystyle{apsrev4-1}
\bibliography{Biblio}

\begin{thebibliography}{41}%
\makeatletter
\providecommand \@ifxundefined [1]{%
 \@ifx{#1\undefined}
}%
\providecommand \@ifnum [1]{%
 \ifnum #1\expandafter \@firstoftwo
 \else \expandafter \@secondoftwo
 \fi
}%
\providecommand \@ifx [1]{%
 \ifx #1\expandafter \@firstoftwo
 \else \expandafter \@secondoftwo
 \fi
}%
\providecommand \natexlab [1]{#1}%
\providecommand \enquote  [1]{``#1''}%
\providecommand \bibnamefont  [1]{#1}%
\providecommand \bibfnamefont [1]{#1}%
\providecommand \citenamefont [1]{#1}%
\providecommand \href@noop [0]{\@secondoftwo}%
\providecommand \href [0]{\begingroup \@sanitize@url \@href}%
\providecommand \@href[1]{\@@startlink{#1}\@@href}%
\providecommand \@@href[1]{\endgroup#1\@@endlink}%
\providecommand \@sanitize@url [0]{\catcode `\\12\catcode `\$12\catcode
  `\&12\catcode `\#12\catcode `\^12\catcode `\_12\catcode `\%12\relax}%
\providecommand \@@startlink[1]{}%
\providecommand \@@endlink[0]{}%
\providecommand \url  [0]{\begingroup\@sanitize@url \@url }%
\providecommand \@url [1]{\endgroup\@href {#1}{\urlprefix }}%
\providecommand \urlprefix  [0]{URL }%
\providecommand \Eprint [0]{\href }%
\providecommand \doibase [0]{http://dx.doi.org/}%
\providecommand \selectlanguage [0]{\@gobble}%
\providecommand \bibinfo  [0]{\@secondoftwo}%
\providecommand \bibfield  [0]{\@secondoftwo}%
\providecommand \translation [1]{[#1]}%
\providecommand \BibitemOpen [0]{}%
\providecommand \bibitemStop [0]{}%
\providecommand \bibitemNoStop [0]{.\EOS\space}%
\providecommand \EOS [0]{\spacefactor3000\relax}%
\providecommand \BibitemShut  [1]{\csname bibitem#1\endcsname}%
\let\auto@bib@innerbib\@empty
\bibitem [{\citenamefont {Schwarz}(2010)}]{schwarz2010mapping}%
  \BibitemOpen
  \bibfield  {author} {\bibinfo {author} {\bibfnamefont {B.}~\bibnamefont
  {Schwarz}},\ }\href@noop {} {\bibfield  {journal} {\bibinfo  {journal}
  {Nature Photonics}\ }\textbf {\bibinfo {volume} {4}},\ \bibinfo {pages} {429}
  (\bibinfo {year} {2010})}\BibitemShut {NoStop}%
\bibitem [{\citenamefont {Velten}\ \emph {et~al.}(2012)\citenamefont {Velten},
  \citenamefont {Willwacher}, \citenamefont {Gupta}, \citenamefont
  {Veeraraghavan}, \citenamefont {Bawendi},\ and\ \citenamefont
  {Raskar}}]{velten2012recovering}%
  \BibitemOpen
  \bibfield  {author} {\bibinfo {author} {\bibfnamefont {A.}~\bibnamefont
  {Velten}}, \bibinfo {author} {\bibfnamefont {T.}~\bibnamefont {Willwacher}},
  \bibinfo {author} {\bibfnamefont {O.}~\bibnamefont {Gupta}}, \bibinfo
  {author} {\bibfnamefont {A.}~\bibnamefont {Veeraraghavan}}, \bibinfo {author}
  {\bibfnamefont {M.~G.}\ \bibnamefont {Bawendi}}, \ and\ \bibinfo {author}
  {\bibfnamefont {R.}~\bibnamefont {Raskar}},\ }\href@noop {} {\bibfield
  {journal} {\bibinfo  {journal} {Nature communications}\ }\textbf {\bibinfo
  {volume} {3}},\ \bibinfo {pages} {1} (\bibinfo {year} {2012})}\BibitemShut
  {NoStop}%
\bibitem [{\citenamefont {Gariepy}\ \emph {et~al.}(2016)\citenamefont
  {Gariepy}, \citenamefont {Tonolini}, \citenamefont {Henderson}, \citenamefont
  {Leach},\ and\ \citenamefont {Faccio}}]{gariepy2016detection}%
  \BibitemOpen
  \bibfield  {author} {\bibinfo {author} {\bibfnamefont {G.}~\bibnamefont
  {Gariepy}}, \bibinfo {author} {\bibfnamefont {F.}~\bibnamefont {Tonolini}},
  \bibinfo {author} {\bibfnamefont {R.}~\bibnamefont {Henderson}}, \bibinfo
  {author} {\bibfnamefont {J.}~\bibnamefont {Leach}}, \ and\ \bibinfo {author}
  {\bibfnamefont {D.}~\bibnamefont {Faccio}},\ }\href@noop {} {\bibfield
  {journal} {\bibinfo  {journal} {Nature Photonics}\ }\textbf {\bibinfo
  {volume} {10}},\ \bibinfo {pages} {23} (\bibinfo {year} {2016})}\BibitemShut
  {NoStop}%
\bibitem [{\citenamefont {O’Toole}\ \emph {et~al.}(2018)\citenamefont
  {O’Toole}, \citenamefont {Lindell},\ and\ \citenamefont
  {Wetzstein}}]{o2018confocal}%
  \BibitemOpen
  \bibfield  {author} {\bibinfo {author} {\bibfnamefont {M.}~\bibnamefont
  {O’Toole}}, \bibinfo {author} {\bibfnamefont {D.~B.}\ \bibnamefont
  {Lindell}}, \ and\ \bibinfo {author} {\bibfnamefont {G.}~\bibnamefont
  {Wetzstein}},\ }\href@noop {} {\bibfield  {journal} {\bibinfo  {journal}
  {Nature}\ }\textbf {\bibinfo {volume} {555}},\ \bibinfo {pages} {338}
  (\bibinfo {year} {2018})}\BibitemShut {NoStop}%
\bibitem [{\citenamefont {Liu}\ \emph {et~al.}(2019{\natexlab{a}})\citenamefont
  {Liu}, \citenamefont {Guill{\'e}n}, \citenamefont {La~Manna}, \citenamefont
  {Nam}, \citenamefont {Reza}, \citenamefont {Le}, \citenamefont {Jarabo},
  \citenamefont {Gutierrez},\ and\ \citenamefont {Velten}}]{liu2019non}%
  \BibitemOpen
  \bibfield  {author} {\bibinfo {author} {\bibfnamefont {X.}~\bibnamefont
  {Liu}}, \bibinfo {author} {\bibfnamefont {I.}~\bibnamefont {Guill{\'e}n}},
  \bibinfo {author} {\bibfnamefont {M.}~\bibnamefont {La~Manna}}, \bibinfo
  {author} {\bibfnamefont {J.~H.}\ \bibnamefont {Nam}}, \bibinfo {author}
  {\bibfnamefont {S.~A.}\ \bibnamefont {Reza}}, \bibinfo {author}
  {\bibfnamefont {T.~H.}\ \bibnamefont {Le}}, \bibinfo {author} {\bibfnamefont
  {A.}~\bibnamefont {Jarabo}}, \bibinfo {author} {\bibfnamefont
  {D.}~\bibnamefont {Gutierrez}}, \ and\ \bibinfo {author} {\bibfnamefont
  {A.}~\bibnamefont {Velten}},\ }\href@noop {} {\bibfield  {journal} {\bibinfo
  {journal} {Nature}\ }\textbf {\bibinfo {volume} {572}},\ \bibinfo {pages}
  {620} (\bibinfo {year} {2019}{\natexlab{a}})}\BibitemShut {NoStop}%
\bibitem [{\citenamefont {Faccio}\ \emph {et~al.}(2020)\citenamefont {Faccio},
  \citenamefont {Velten},\ and\ \citenamefont {Wetzstein}}]{faccio2020non}%
  \BibitemOpen
  \bibfield  {author} {\bibinfo {author} {\bibfnamefont {D.}~\bibnamefont
  {Faccio}}, \bibinfo {author} {\bibfnamefont {A.}~\bibnamefont {Velten}}, \
  and\ \bibinfo {author} {\bibfnamefont {G.}~\bibnamefont {Wetzstein}},\
  }\href@noop {} {\bibfield  {journal} {\bibinfo  {journal} {Nature Reviews
  Physics}\ }\textbf {\bibinfo {volume} {2}},\ \bibinfo {pages} {318} (\bibinfo
  {year} {2020})}\BibitemShut {NoStop}%
\bibitem [{\citenamefont {Lyons}\ \emph {et~al.}(2019)\citenamefont {Lyons},
  \citenamefont {Tonolini}, \citenamefont {Boccolini}, \citenamefont {Repetti},
  \citenamefont {Henderson}, \citenamefont {Wiaux},\ and\ \citenamefont
  {Faccio}}]{lyons2019computational}%
  \BibitemOpen
  \bibfield  {author} {\bibinfo {author} {\bibfnamefont {A.}~\bibnamefont
  {Lyons}}, \bibinfo {author} {\bibfnamefont {F.}~\bibnamefont {Tonolini}},
  \bibinfo {author} {\bibfnamefont {A.}~\bibnamefont {Boccolini}}, \bibinfo
  {author} {\bibfnamefont {A.}~\bibnamefont {Repetti}}, \bibinfo {author}
  {\bibfnamefont {R.}~\bibnamefont {Henderson}}, \bibinfo {author}
  {\bibfnamefont {Y.}~\bibnamefont {Wiaux}}, \ and\ \bibinfo {author}
  {\bibfnamefont {D.}~\bibnamefont {Faccio}},\ }\href@noop {} {\bibfield
  {journal} {\bibinfo  {journal} {Nature Photonics}\ }\textbf {\bibinfo
  {volume} {13}},\ \bibinfo {pages} {575} (\bibinfo {year} {2019})}\BibitemShut
  {NoStop}%
\bibitem [{\citenamefont {Bruschini}\ \emph {et~al.}(2019)\citenamefont
  {Bruschini}, \citenamefont {Homulle}, \citenamefont {Antolovic},
  \citenamefont {Burri},\ and\ \citenamefont {Charbon}}]{bruschini2019single}%
  \BibitemOpen
  \bibfield  {author} {\bibinfo {author} {\bibfnamefont {C.}~\bibnamefont
  {Bruschini}}, \bibinfo {author} {\bibfnamefont {H.}~\bibnamefont {Homulle}},
  \bibinfo {author} {\bibfnamefont {I.~M.}\ \bibnamefont {Antolovic}}, \bibinfo
  {author} {\bibfnamefont {S.}~\bibnamefont {Burri}}, \ and\ \bibinfo {author}
  {\bibfnamefont {E.}~\bibnamefont {Charbon}},\ }\href@noop {} {\bibfield
  {journal} {\bibinfo  {journal} {Light: Science \& Applications}\ }\textbf
  {\bibinfo {volume} {8}},\ \bibinfo {pages} {1} (\bibinfo {year}
  {2019})}\BibitemShut {NoStop}%
\bibitem [{\citenamefont {Sun}\ \emph {et~al.}(2016)\citenamefont {Sun},
  \citenamefont {Edgar}, \citenamefont {Gibson}, \citenamefont {Sun},
  \citenamefont {Radwell}, \citenamefont {Lamb},\ and\ \citenamefont
  {Padgett}}]{sun2016single}%
  \BibitemOpen
  \bibfield  {author} {\bibinfo {author} {\bibfnamefont {M.-J.}\ \bibnamefont
  {Sun}}, \bibinfo {author} {\bibfnamefont {M.~P.}\ \bibnamefont {Edgar}},
  \bibinfo {author} {\bibfnamefont {G.~M.}\ \bibnamefont {Gibson}}, \bibinfo
  {author} {\bibfnamefont {B.}~\bibnamefont {Sun}}, \bibinfo {author}
  {\bibfnamefont {N.}~\bibnamefont {Radwell}}, \bibinfo {author} {\bibfnamefont
  {R.}~\bibnamefont {Lamb}}, \ and\ \bibinfo {author} {\bibfnamefont {M.~J.}\
  \bibnamefont {Padgett}},\ }\href@noop {} {\bibfield  {journal} {\bibinfo
  {journal} {Nature communications}\ }\textbf {\bibinfo {volume} {7}},\
  \bibinfo {pages} {1} (\bibinfo {year} {2016})}\BibitemShut {NoStop}%
\bibitem [{\citenamefont {Shin}\ \emph {et~al.}(2016)\citenamefont {Shin},
  \citenamefont {Xu}, \citenamefont {Venkatraman}, \citenamefont {Lussana},
  \citenamefont {Villa}, \citenamefont {Zappa}, \citenamefont {Goyal},
  \citenamefont {Wong},\ and\ \citenamefont {Shapiro}}]{shin2016photon}%
  \BibitemOpen
  \bibfield  {author} {\bibinfo {author} {\bibfnamefont {D.}~\bibnamefont
  {Shin}}, \bibinfo {author} {\bibfnamefont {F.}~\bibnamefont {Xu}}, \bibinfo
  {author} {\bibfnamefont {D.}~\bibnamefont {Venkatraman}}, \bibinfo {author}
  {\bibfnamefont {R.}~\bibnamefont {Lussana}}, \bibinfo {author} {\bibfnamefont
  {F.}~\bibnamefont {Villa}}, \bibinfo {author} {\bibfnamefont
  {F.}~\bibnamefont {Zappa}}, \bibinfo {author} {\bibfnamefont {V.~K.}\
  \bibnamefont {Goyal}}, \bibinfo {author} {\bibfnamefont {F.~N.}\ \bibnamefont
  {Wong}}, \ and\ \bibinfo {author} {\bibfnamefont {J.~H.}\ \bibnamefont
  {Shapiro}},\ }\href@noop {} {\bibfield  {journal} {\bibinfo  {journal}
  {Nature communications}\ }\textbf {\bibinfo {volume} {7}},\ \bibinfo {pages}
  {1} (\bibinfo {year} {2016})}\BibitemShut {NoStop}%
\bibitem [{\citenamefont {Tachella}\ \emph {et~al.}(2019)\citenamefont
  {Tachella}, \citenamefont {Altmann}, \citenamefont {Mellado}, \citenamefont
  {McCarthy}, \citenamefont {Tobin}, \citenamefont {Buller}, \citenamefont
  {Tourneret},\ and\ \citenamefont {McLaughlin}}]{tachella2019real}%
  \BibitemOpen
  \bibfield  {author} {\bibinfo {author} {\bibfnamefont {J.}~\bibnamefont
  {Tachella}}, \bibinfo {author} {\bibfnamefont {Y.}~\bibnamefont {Altmann}},
  \bibinfo {author} {\bibfnamefont {N.}~\bibnamefont {Mellado}}, \bibinfo
  {author} {\bibfnamefont {A.}~\bibnamefont {McCarthy}}, \bibinfo {author}
  {\bibfnamefont {R.}~\bibnamefont {Tobin}}, \bibinfo {author} {\bibfnamefont
  {G.~S.}\ \bibnamefont {Buller}}, \bibinfo {author} {\bibfnamefont {J.-Y.}\
  \bibnamefont {Tourneret}}, \ and\ \bibinfo {author} {\bibfnamefont
  {S.}~\bibnamefont {McLaughlin}},\ }\href@noop {} {\bibfield  {journal}
  {\bibinfo  {journal} {Nature communications}\ }\textbf {\bibinfo {volume}
  {10}},\ \bibinfo {pages} {1} (\bibinfo {year} {2019})}\BibitemShut {NoStop}%
\bibitem [{\citenamefont {Zhang}\ \emph {et~al.}(2018)\citenamefont {Zhang},
  \citenamefont {Lindner}, \citenamefont {Antolovi{\'c}}, \citenamefont
  {Pavia}, \citenamefont {Wolf},\ and\ \citenamefont {Charbon}}]{zhang201830}%
  \BibitemOpen
  \bibfield  {author} {\bibinfo {author} {\bibfnamefont {C.}~\bibnamefont
  {Zhang}}, \bibinfo {author} {\bibfnamefont {S.}~\bibnamefont {Lindner}},
  \bibinfo {author} {\bibfnamefont {I.~M.}\ \bibnamefont {Antolovi{\'c}}},
  \bibinfo {author} {\bibfnamefont {J.~M.}\ \bibnamefont {Pavia}}, \bibinfo
  {author} {\bibfnamefont {M.}~\bibnamefont {Wolf}}, \ and\ \bibinfo {author}
  {\bibfnamefont {E.}~\bibnamefont {Charbon}},\ }\href@noop {} {\bibfield
  {journal} {\bibinfo  {journal} {IEEE Journal of Solid-State Circuits}\
  }\textbf {\bibinfo {volume} {54}},\ \bibinfo {pages} {1137} (\bibinfo {year}
  {2018})}\BibitemShut {NoStop}%
\bibitem [{\citenamefont {{Ronchini Ximenes}}\ \emph
  {et~al.}(2019)\citenamefont {{Ronchini Ximenes}}, \citenamefont
  {{Padmanabhan}}, \citenamefont {{Lee}}, \citenamefont {{Yamashita}},
  \citenamefont {{Yaung}},\ and\ \citenamefont
  {{Charbon}}}]{ximenes_JSSC_2019}%
  \BibitemOpen
  \bibfield  {author} {\bibinfo {author} {\bibfnamefont {A.}~\bibnamefont
  {{Ronchini Ximenes}}}, \bibinfo {author} {\bibfnamefont {P.}~\bibnamefont
  {{Padmanabhan}}}, \bibinfo {author} {\bibfnamefont {M.}~\bibnamefont
  {{Lee}}}, \bibinfo {author} {\bibfnamefont {Y.}~\bibnamefont {{Yamashita}}},
  \bibinfo {author} {\bibfnamefont {D.}~\bibnamefont {{Yaung}}}, \ and\
  \bibinfo {author} {\bibfnamefont {E.}~\bibnamefont {{Charbon}}},\ }\href
  {\doibase 10.1109/JSSC.2019.2938412} {\bibfield  {journal} {\bibinfo
  {journal} {IEEE Journal of Solid-State Circuits}\ }\textbf {\bibinfo {volume}
  {54}},\ \bibinfo {pages} {3203} (\bibinfo {year} {2019})}\BibitemShut
  {NoStop}%
\bibitem [{\citenamefont {{Seo}}\ \emph {et~al.}(2021)\citenamefont {{Seo}},
  \citenamefont {{Yoon}}, \citenamefont {{Kim}}, \citenamefont {{Kim}},
  \citenamefont {{Kim}}, \citenamefont {{Chun}},\ and\ \citenamefont
  {{Choi}}}]{choi_JSSC_2021}%
  \BibitemOpen
  \bibfield  {author} {\bibinfo {author} {\bibfnamefont {H.}~\bibnamefont
  {{Seo}}}, \bibinfo {author} {\bibfnamefont {H.}~\bibnamefont {{Yoon}}},
  \bibinfo {author} {\bibfnamefont {D.}~\bibnamefont {{Kim}}}, \bibinfo
  {author} {\bibfnamefont {J.}~\bibnamefont {{Kim}}}, \bibinfo {author}
  {\bibfnamefont {S.~J.}\ \bibnamefont {{Kim}}}, \bibinfo {author}
  {\bibfnamefont {J.~H.}\ \bibnamefont {{Chun}}}, \ and\ \bibinfo {author}
  {\bibfnamefont {J.}~\bibnamefont {{Choi}}},\ }\href {\doibase
  10.1109/JSSC.2020.3048074} {\bibfield  {journal} {\bibinfo  {journal} {IEEE
  Journal of Solid-State Circuits}\ ,\ \bibinfo {pages} {1}} (\bibinfo {year}
  {2021})}\BibitemShut {NoStop}%
\bibitem [{\citenamefont {Ulku}\ \emph {et~al.}(2018)\citenamefont {Ulku},
  \citenamefont {Bruschini}, \citenamefont {Antolovi{\'c}}, \citenamefont
  {Kuo}, \citenamefont {Ankri}, \citenamefont {Weiss}, \citenamefont
  {Michalet},\ and\ \citenamefont {Charbon}}]{ulku2018512}%
  \BibitemOpen
  \bibfield  {author} {\bibinfo {author} {\bibfnamefont {A.~C.}\ \bibnamefont
  {Ulku}}, \bibinfo {author} {\bibfnamefont {C.}~\bibnamefont {Bruschini}},
  \bibinfo {author} {\bibfnamefont {I.~M.}\ \bibnamefont {Antolovi{\'c}}},
  \bibinfo {author} {\bibfnamefont {Y.}~\bibnamefont {Kuo}}, \bibinfo {author}
  {\bibfnamefont {R.}~\bibnamefont {Ankri}}, \bibinfo {author} {\bibfnamefont
  {S.}~\bibnamefont {Weiss}}, \bibinfo {author} {\bibfnamefont
  {X.}~\bibnamefont {Michalet}}, \ and\ \bibinfo {author} {\bibfnamefont
  {E.}~\bibnamefont {Charbon}},\ }\href@noop {} {\bibfield  {journal} {\bibinfo
   {journal} {IEEE Journal of Selected Topics in Quantum Electronics}\ }\textbf
  {\bibinfo {volume} {25}},\ \bibinfo {pages} {1} (\bibinfo {year}
  {2018})}\BibitemShut {NoStop}%
\bibitem [{\citenamefont {Morimoto}\ \emph {et~al.}(2020)\citenamefont
  {Morimoto}, \citenamefont {Ardelean}, \citenamefont {Wu}, \citenamefont
  {Ulku}, \citenamefont {Antolovic}, \citenamefont {Bruschini},\ and\
  \citenamefont {Charbon}}]{morimoto2020megapixel}%
  \BibitemOpen
  \bibfield  {author} {\bibinfo {author} {\bibfnamefont {K.}~\bibnamefont
  {Morimoto}}, \bibinfo {author} {\bibfnamefont {A.}~\bibnamefont {Ardelean}},
  \bibinfo {author} {\bibfnamefont {M.-L.}\ \bibnamefont {Wu}}, \bibinfo
  {author} {\bibfnamefont {A.~C.}\ \bibnamefont {Ulku}}, \bibinfo {author}
  {\bibfnamefont {I.~M.}\ \bibnamefont {Antolovic}}, \bibinfo {author}
  {\bibfnamefont {C.}~\bibnamefont {Bruschini}}, \ and\ \bibinfo {author}
  {\bibfnamefont {E.}~\bibnamefont {Charbon}},\ }\href@noop {} {\bibfield
  {journal} {\bibinfo  {journal} {Optica}\ }\textbf {\bibinfo {volume} {7}},\
  \bibinfo {pages} {346} (\bibinfo {year} {2020})}\BibitemShut {NoStop}%
\bibitem [{\citenamefont {Ren}\ \emph {et~al.}(2018)\citenamefont {Ren},
  \citenamefont {Connolly}, \citenamefont {Halimi}, \citenamefont {Altmann},
  \citenamefont {McLaughlin}, \citenamefont {Gyongy}, \citenamefont
  {Henderson},\ and\ \citenamefont {Buller}}]{ren2018high}%
  \BibitemOpen
  \bibfield  {author} {\bibinfo {author} {\bibfnamefont {X.}~\bibnamefont
  {Ren}}, \bibinfo {author} {\bibfnamefont {P.~W.}\ \bibnamefont {Connolly}},
  \bibinfo {author} {\bibfnamefont {A.}~\bibnamefont {Halimi}}, \bibinfo
  {author} {\bibfnamefont {Y.}~\bibnamefont {Altmann}}, \bibinfo {author}
  {\bibfnamefont {S.}~\bibnamefont {McLaughlin}}, \bibinfo {author}
  {\bibfnamefont {I.}~\bibnamefont {Gyongy}}, \bibinfo {author} {\bibfnamefont
  {R.~K.}\ \bibnamefont {Henderson}}, \ and\ \bibinfo {author} {\bibfnamefont
  {G.~S.}\ \bibnamefont {Buller}},\ }\href@noop {} {\bibfield  {journal}
  {\bibinfo  {journal} {Optics express}\ }\textbf {\bibinfo {volume} {26}},\
  \bibinfo {pages} {5541} (\bibinfo {year} {2018})}\BibitemShut {NoStop}%
\bibitem [{\citenamefont {Chan}\ \emph {et~al.}(2019)\citenamefont {Chan},
  \citenamefont {Halimi}, \citenamefont {Zhu}, \citenamefont {Gyongy},
  \citenamefont {Henderson}, \citenamefont {Bowman}, \citenamefont
  {McLaughlin}, \citenamefont {Buller},\ and\ \citenamefont
  {Leach}}]{chan2019long}%
  \BibitemOpen
  \bibfield  {author} {\bibinfo {author} {\bibfnamefont {S.}~\bibnamefont
  {Chan}}, \bibinfo {author} {\bibfnamefont {A.}~\bibnamefont {Halimi}},
  \bibinfo {author} {\bibfnamefont {F.}~\bibnamefont {Zhu}}, \bibinfo {author}
  {\bibfnamefont {I.}~\bibnamefont {Gyongy}}, \bibinfo {author} {\bibfnamefont
  {R.~K.}\ \bibnamefont {Henderson}}, \bibinfo {author} {\bibfnamefont
  {R.}~\bibnamefont {Bowman}}, \bibinfo {author} {\bibfnamefont
  {S.}~\bibnamefont {McLaughlin}}, \bibinfo {author} {\bibfnamefont {G.~S.}\
  \bibnamefont {Buller}}, \ and\ \bibinfo {author} {\bibfnamefont
  {J.}~\bibnamefont {Leach}},\ }\href@noop {} {\bibfield  {journal} {\bibinfo
  {journal} {Scientific reports}\ }\textbf {\bibinfo {volume} {9}},\ \bibinfo
  {pages} {1} (\bibinfo {year} {2019})}\BibitemShut {NoStop}%
\bibitem [{\citenamefont {{Niclass}}\ \emph {et~al.}(2013)\citenamefont
  {{Niclass}}, \citenamefont {{Soga}}, \citenamefont {{Matsubara}},
  \citenamefont {{Kato}},\ and\ \citenamefont {{Kagami}}}]{niclass_JSSC_2013}%
  \BibitemOpen
  \bibfield  {author} {\bibinfo {author} {\bibfnamefont {C.}~\bibnamefont
  {{Niclass}}}, \bibinfo {author} {\bibfnamefont {M.}~\bibnamefont {{Soga}}},
  \bibinfo {author} {\bibfnamefont {H.}~\bibnamefont {{Matsubara}}}, \bibinfo
  {author} {\bibfnamefont {S.}~\bibnamefont {{Kato}}}, \ and\ \bibinfo {author}
  {\bibfnamefont {M.}~\bibnamefont {{Kagami}}},\ }\href {\doibase
  10.1109/JSSC.2012.2227607} {\bibfield  {journal} {\bibinfo  {journal} {IEEE
  Journal of Solid-State Circuits}\ }\textbf {\bibinfo {volume} {48}},\
  \bibinfo {pages} {559} (\bibinfo {year} {2013})}\BibitemShut {NoStop}%
\bibitem [{\citenamefont {Lloyd}(2008)}]{lloyd2008enhanced}%
  \BibitemOpen
  \bibfield  {author} {\bibinfo {author} {\bibfnamefont {S.}~\bibnamefont
  {Lloyd}},\ }\href@noop {} {\bibfield  {journal} {\bibinfo  {journal}
  {Science}\ }\textbf {\bibinfo {volume} {321}},\ \bibinfo {pages} {1463}
  (\bibinfo {year} {2008})}\BibitemShut {NoStop}%
\bibitem [{\citenamefont {Lopaeva}\ \emph {et~al.}(2013)\citenamefont
  {Lopaeva}, \citenamefont {Berchera}, \citenamefont {Degiovanni},
  \citenamefont {Olivares}, \citenamefont {Brida},\ and\ \citenamefont
  {Genovese}}]{lopaeva2013experimental}%
  \BibitemOpen
  \bibfield  {author} {\bibinfo {author} {\bibfnamefont {E.}~\bibnamefont
  {Lopaeva}}, \bibinfo {author} {\bibfnamefont {I.~R.}\ \bibnamefont
  {Berchera}}, \bibinfo {author} {\bibfnamefont {I.~P.}\ \bibnamefont
  {Degiovanni}}, \bibinfo {author} {\bibfnamefont {S.}~\bibnamefont
  {Olivares}}, \bibinfo {author} {\bibfnamefont {G.}~\bibnamefont {Brida}}, \
  and\ \bibinfo {author} {\bibfnamefont {M.}~\bibnamefont {Genovese}},\
  }\href@noop {} {\bibfield  {journal} {\bibinfo  {journal} {Physical review
  letters}\ }\textbf {\bibinfo {volume} {110}},\ \bibinfo {pages} {153603}
  (\bibinfo {year} {2013})}\BibitemShut {NoStop}%
\bibitem [{\citenamefont {Zhang}\ \emph {et~al.}(2020)\citenamefont {Zhang},
  \citenamefont {England}, \citenamefont {Nomerotski}, \citenamefont {Svihra},
  \citenamefont {Ferrante}, \citenamefont {Hockett},\ and\ \citenamefont
  {Sussman}}]{zhang2020multidimensional}%
  \BibitemOpen
  \bibfield  {author} {\bibinfo {author} {\bibfnamefont {Y.}~\bibnamefont
  {Zhang}}, \bibinfo {author} {\bibfnamefont {D.}~\bibnamefont {England}},
  \bibinfo {author} {\bibfnamefont {A.}~\bibnamefont {Nomerotski}}, \bibinfo
  {author} {\bibfnamefont {P.}~\bibnamefont {Svihra}}, \bibinfo {author}
  {\bibfnamefont {S.}~\bibnamefont {Ferrante}}, \bibinfo {author}
  {\bibfnamefont {P.}~\bibnamefont {Hockett}}, \ and\ \bibinfo {author}
  {\bibfnamefont {B.}~\bibnamefont {Sussman}},\ }\href@noop {} {\bibfield
  {journal} {\bibinfo  {journal} {Physical Review A}\ }\textbf {\bibinfo
  {volume} {101}},\ \bibinfo {pages} {053808} (\bibinfo {year}
  {2020})}\BibitemShut {NoStop}%
\bibitem [{\citenamefont {England}\ \emph {et~al.}(2019)\citenamefont
  {England}, \citenamefont {Balaji},\ and\ \citenamefont
  {Sussman}}]{england2019quantum}%
  \BibitemOpen
  \bibfield  {author} {\bibinfo {author} {\bibfnamefont {D.~G.}\ \bibnamefont
  {England}}, \bibinfo {author} {\bibfnamefont {B.}~\bibnamefont {Balaji}}, \
  and\ \bibinfo {author} {\bibfnamefont {B.~J.}\ \bibnamefont {Sussman}},\
  }\href@noop {} {\bibfield  {journal} {\bibinfo  {journal} {Physical Review
  A}\ }\textbf {\bibinfo {volume} {99}},\ \bibinfo {pages} {023828} (\bibinfo
  {year} {2019})}\BibitemShut {NoStop}%
\bibitem [{\citenamefont {Defienne}\ \emph {et~al.}(2019)\citenamefont
  {Defienne}, \citenamefont {Reichert}, \citenamefont {Fleischer},\ and\
  \citenamefont {Faccio}}]{defienne2019quantum}%
  \BibitemOpen
  \bibfield  {author} {\bibinfo {author} {\bibfnamefont {H.}~\bibnamefont
  {Defienne}}, \bibinfo {author} {\bibfnamefont {M.}~\bibnamefont {Reichert}},
  \bibinfo {author} {\bibfnamefont {J.~W.}\ \bibnamefont {Fleischer}}, \ and\
  \bibinfo {author} {\bibfnamefont {D.}~\bibnamefont {Faccio}},\ }\href@noop {}
  {\bibfield  {journal} {\bibinfo  {journal} {Science advances}\ }\textbf
  {\bibinfo {volume} {5}},\ \bibinfo {pages} {eaax0307} (\bibinfo {year}
  {2019})}\BibitemShut {NoStop}%
\bibitem [{\citenamefont {Gregory}\ \emph {et~al.}(2020)\citenamefont
  {Gregory}, \citenamefont {Moreau}, \citenamefont {Toninelli},\ and\
  \citenamefont {Padgett}}]{gregory2020imaging}%
  \BibitemOpen
  \bibfield  {author} {\bibinfo {author} {\bibfnamefont {T.}~\bibnamefont
  {Gregory}}, \bibinfo {author} {\bibfnamefont {P.-A.}\ \bibnamefont {Moreau}},
  \bibinfo {author} {\bibfnamefont {E.}~\bibnamefont {Toninelli}}, \ and\
  \bibinfo {author} {\bibfnamefont {M.~J.}\ \bibnamefont {Padgett}},\
  }\href@noop {} {\bibfield  {journal} {\bibinfo  {journal} {Science advances}\
  }\textbf {\bibinfo {volume} {6}},\ \bibinfo {pages} {eaay2652} (\bibinfo
  {year} {2020})}\BibitemShut {NoStop}%
\bibitem [{\citenamefont {Moreau}\ \emph {et~al.}(2012)\citenamefont {Moreau},
  \citenamefont {Mougin-Sisini}, \citenamefont {Devaux},\ and\ \citenamefont
  {Lantz}}]{moreau2012realization}%
  \BibitemOpen
  \bibfield  {author} {\bibinfo {author} {\bibfnamefont {P.-A.}\ \bibnamefont
  {Moreau}}, \bibinfo {author} {\bibfnamefont {J.}~\bibnamefont
  {Mougin-Sisini}}, \bibinfo {author} {\bibfnamefont {F.}~\bibnamefont
  {Devaux}}, \ and\ \bibinfo {author} {\bibfnamefont {E.}~\bibnamefont
  {Lantz}},\ }\href@noop {} {\bibfield  {journal} {\bibinfo  {journal}
  {Physical Review A}\ }\textbf {\bibinfo {volume} {86}},\ \bibinfo {pages}
  {010101} (\bibinfo {year} {2012})}\BibitemShut {NoStop}%
\bibitem [{\citenamefont {Edgar}\ \emph {et~al.}(2012)\citenamefont {Edgar},
  \citenamefont {Tasca}, \citenamefont {Izdebski}, \citenamefont {Warburton},
  \citenamefont {Leach}, \citenamefont {Agnew}, \citenamefont {Buller},
  \citenamefont {Boyd},\ and\ \citenamefont {Padgett}}]{edgar2012imaging}%
  \BibitemOpen
  \bibfield  {author} {\bibinfo {author} {\bibfnamefont {M.~P.}\ \bibnamefont
  {Edgar}}, \bibinfo {author} {\bibfnamefont {D.~S.}\ \bibnamefont {Tasca}},
  \bibinfo {author} {\bibfnamefont {F.}~\bibnamefont {Izdebski}}, \bibinfo
  {author} {\bibfnamefont {R.~E.}\ \bibnamefont {Warburton}}, \bibinfo {author}
  {\bibfnamefont {J.}~\bibnamefont {Leach}}, \bibinfo {author} {\bibfnamefont
  {M.}~\bibnamefont {Agnew}}, \bibinfo {author} {\bibfnamefont {G.~S.}\
  \bibnamefont {Buller}}, \bibinfo {author} {\bibfnamefont {R.~W.}\
  \bibnamefont {Boyd}}, \ and\ \bibinfo {author} {\bibfnamefont {M.~J.}\
  \bibnamefont {Padgett}},\ }\href@noop {} {\bibfield  {journal} {\bibinfo
  {journal} {Nature communications}\ }\textbf {\bibinfo {volume} {3}},\
  \bibinfo {pages} {1} (\bibinfo {year} {2012})}\BibitemShut {NoStop}%
\bibitem [{\citenamefont {Defienne}\ \emph {et~al.}(2018)\citenamefont
  {Defienne}, \citenamefont {Reichert},\ and\ \citenamefont
  {Fleischer}}]{defienne2018general}%
  \BibitemOpen
  \bibfield  {author} {\bibinfo {author} {\bibfnamefont {H.}~\bibnamefont
  {Defienne}}, \bibinfo {author} {\bibfnamefont {M.}~\bibnamefont {Reichert}},
  \ and\ \bibinfo {author} {\bibfnamefont {J.~W.}\ \bibnamefont {Fleischer}},\
  }\href@noop {} {\bibfield  {journal} {\bibinfo  {journal} {Physical review
  letters}\ }\textbf {\bibinfo {volume} {120}},\ \bibinfo {pages} {203604}
  (\bibinfo {year} {2018})}\BibitemShut {NoStop}%
\bibitem [{\citenamefont {Chrapkiewicz}\ \emph {et~al.}(2014)\citenamefont
  {Chrapkiewicz}, \citenamefont {Wasilewski},\ and\ \citenamefont
  {Banaszek}}]{chrapkiewicz2014high}%
  \BibitemOpen
  \bibfield  {author} {\bibinfo {author} {\bibfnamefont {R.}~\bibnamefont
  {Chrapkiewicz}}, \bibinfo {author} {\bibfnamefont {W.}~\bibnamefont
  {Wasilewski}}, \ and\ \bibinfo {author} {\bibfnamefont {K.}~\bibnamefont
  {Banaszek}},\ }\href@noop {} {\bibfield  {journal} {\bibinfo  {journal}
  {Optics letters}\ }\textbf {\bibinfo {volume} {39}},\ \bibinfo {pages} {5090}
  (\bibinfo {year} {2014})}\BibitemShut {NoStop}%
\bibitem [{\citenamefont {Chrapkiewicz}\ \emph {et~al.}(2016)\citenamefont
  {Chrapkiewicz}, \citenamefont {Jachura}, \citenamefont {Banaszek},\ and\
  \citenamefont {Wasilewski}}]{chrapkiewicz2016hologram}%
  \BibitemOpen
  \bibfield  {author} {\bibinfo {author} {\bibfnamefont {R.}~\bibnamefont
  {Chrapkiewicz}}, \bibinfo {author} {\bibfnamefont {M.}~\bibnamefont
  {Jachura}}, \bibinfo {author} {\bibfnamefont {K.}~\bibnamefont {Banaszek}}, \
  and\ \bibinfo {author} {\bibfnamefont {W.}~\bibnamefont {Wasilewski}},\
  }\href@noop {} {\bibfield  {journal} {\bibinfo  {journal} {Nature Photonics}\
  }\textbf {\bibinfo {volume} {10}},\ \bibinfo {pages} {576} (\bibinfo {year}
  {2016})}\BibitemShut {NoStop}%
\bibitem [{\citenamefont {Eckmann}\ \emph {et~al.}(2020)\citenamefont
  {Eckmann}, \citenamefont {Bessire}, \citenamefont {Untern{\"a}hrer},
  \citenamefont {Gasparini}, \citenamefont {Perenzoni},\ and\ \citenamefont
  {Stefanov}}]{eckmann2020characterization}%
  \BibitemOpen
  \bibfield  {author} {\bibinfo {author} {\bibfnamefont {B.}~\bibnamefont
  {Eckmann}}, \bibinfo {author} {\bibfnamefont {B.}~\bibnamefont {Bessire}},
  \bibinfo {author} {\bibfnamefont {M.}~\bibnamefont {Untern{\"a}hrer}},
  \bibinfo {author} {\bibfnamefont {L.}~\bibnamefont {Gasparini}}, \bibinfo
  {author} {\bibfnamefont {M.}~\bibnamefont {Perenzoni}}, \ and\ \bibinfo
  {author} {\bibfnamefont {A.}~\bibnamefont {Stefanov}},\ }\href@noop {}
  {\bibfield  {journal} {\bibinfo  {journal} {Optics Express}\ }\textbf
  {\bibinfo {volume} {28}},\ \bibinfo {pages} {31553} (\bibinfo {year}
  {2020})}\BibitemShut {NoStop}%
\bibitem [{\citenamefont {Ianzano}\ \emph {et~al.}(2020)\citenamefont
  {Ianzano}, \citenamefont {Svihra}, \citenamefont {Flament}, \citenamefont
  {Hardy}, \citenamefont {Cui}, \citenamefont {Nomerotski},\ and\ \citenamefont
  {Figueroa}}]{ianzano2020fast}%
  \BibitemOpen
  \bibfield  {author} {\bibinfo {author} {\bibfnamefont {C.}~\bibnamefont
  {Ianzano}}, \bibinfo {author} {\bibfnamefont {P.}~\bibnamefont {Svihra}},
  \bibinfo {author} {\bibfnamefont {M.}~\bibnamefont {Flament}}, \bibinfo
  {author} {\bibfnamefont {A.}~\bibnamefont {Hardy}}, \bibinfo {author}
  {\bibfnamefont {G.}~\bibnamefont {Cui}}, \bibinfo {author} {\bibfnamefont
  {A.}~\bibnamefont {Nomerotski}}, \ and\ \bibinfo {author} {\bibfnamefont
  {E.}~\bibnamefont {Figueroa}},\ }\href@noop {} {\bibfield  {journal}
  {\bibinfo  {journal} {Scientific reports}\ }\textbf {\bibinfo {volume}
  {10}},\ \bibinfo {pages} {1} (\bibinfo {year} {2020})}\BibitemShut {NoStop}%
\bibitem [{\citenamefont {Ndagano}\ \emph {et~al.}(2020)\citenamefont
  {Ndagano}, \citenamefont {Defienne}, \citenamefont {Lyons}, \citenamefont
  {Starshynov}, \citenamefont {Villa}, \citenamefont {Tisa},\ and\
  \citenamefont {Faccio}}]{ndagano2020imaging}%
  \BibitemOpen
  \bibfield  {author} {\bibinfo {author} {\bibfnamefont {B.}~\bibnamefont
  {Ndagano}}, \bibinfo {author} {\bibfnamefont {H.}~\bibnamefont {Defienne}},
  \bibinfo {author} {\bibfnamefont {A.}~\bibnamefont {Lyons}}, \bibinfo
  {author} {\bibfnamefont {I.}~\bibnamefont {Starshynov}}, \bibinfo {author}
  {\bibfnamefont {F.}~\bibnamefont {Villa}}, \bibinfo {author} {\bibfnamefont
  {S.}~\bibnamefont {Tisa}}, \ and\ \bibinfo {author} {\bibfnamefont
  {D.}~\bibnamefont {Faccio}},\ }\href@noop {} {\bibfield  {journal} {\bibinfo
  {journal} {npj Quantum Information}\ }\textbf {\bibinfo {volume} {6}},\
  \bibinfo {pages} {1} (\bibinfo {year} {2020})}\BibitemShut {NoStop}%
\bibitem [{\citenamefont {Defienne}\ \emph {et~al.}(2021)\citenamefont
  {Defienne}, \citenamefont {Zhao}, \citenamefont {Charbon},\ and\
  \citenamefont {Faccio}}]{defienne2021full}%
  \BibitemOpen
  \bibfield  {author} {\bibinfo {author} {\bibfnamefont {H.}~\bibnamefont
  {Defienne}}, \bibinfo {author} {\bibfnamefont {J.}~\bibnamefont {Zhao}},
  \bibinfo {author} {\bibfnamefont {E.}~\bibnamefont {Charbon}}, \ and\
  \bibinfo {author} {\bibfnamefont {D.}~\bibnamefont {Faccio}},\ }\href@noop {}
  {\bibfield  {journal} {\bibinfo  {journal} {Physical Review A}\ }\textbf
  {\bibinfo {volume} {103}},\ \bibinfo {pages} {042608} (\bibinfo {year}
  {2021})}\BibitemShut {NoStop}%
\bibitem [{\citenamefont {Liu}\ \emph {et~al.}(2019{\natexlab{b}})\citenamefont
  {Liu}, \citenamefont {Giovannini}, \citenamefont {He}, \citenamefont
  {England}, \citenamefont {Sussman}, \citenamefont {Balaji},\ and\
  \citenamefont {Helmy}}]{liu2019enhancing}%
  \BibitemOpen
  \bibfield  {author} {\bibinfo {author} {\bibfnamefont {H.}~\bibnamefont
  {Liu}}, \bibinfo {author} {\bibfnamefont {D.}~\bibnamefont {Giovannini}},
  \bibinfo {author} {\bibfnamefont {H.}~\bibnamefont {He}}, \bibinfo {author}
  {\bibfnamefont {D.}~\bibnamefont {England}}, \bibinfo {author} {\bibfnamefont
  {B.~J.}\ \bibnamefont {Sussman}}, \bibinfo {author} {\bibfnamefont
  {B.}~\bibnamefont {Balaji}}, \ and\ \bibinfo {author} {\bibfnamefont {A.~S.}\
  \bibnamefont {Helmy}},\ }\href@noop {} {\bibfield  {journal} {\bibinfo
  {journal} {Optica}\ }\textbf {\bibinfo {volume} {6}},\ \bibinfo {pages}
  {1349} (\bibinfo {year} {2019}{\natexlab{b}})}\BibitemShut {NoStop}%
\bibitem [{\citenamefont {Frick}\ \emph {et~al.}(2020)\citenamefont {Frick},
  \citenamefont {McMillan},\ and\ \citenamefont {Rarity}}]{frick2020quantum}%
  \BibitemOpen
  \bibfield  {author} {\bibinfo {author} {\bibfnamefont {S.}~\bibnamefont
  {Frick}}, \bibinfo {author} {\bibfnamefont {A.}~\bibnamefont {McMillan}}, \
  and\ \bibinfo {author} {\bibfnamefont {J.}~\bibnamefont {Rarity}},\
  }\href@noop {} {\bibfield  {journal} {\bibinfo  {journal} {Optics Express}\
  }\textbf {\bibinfo {volume} {28}},\ \bibinfo {pages} {37118} (\bibinfo {year}
  {2020})}\BibitemShut {NoStop}%
\bibitem [{\citenamefont {Ren}\ \emph {et~al.}(2020)\citenamefont {Ren},
  \citenamefont {Frick}, \citenamefont {McMillan}, \citenamefont {Chen},
  \citenamefont {Halimi}, \citenamefont {Connolly}, \citenamefont {Joshi},
  \citenamefont {Mclaughlin}, \citenamefont {Rarity}, \citenamefont {Matthews}
  \emph {et~al.}}]{ren2020time}%
  \BibitemOpen
  \bibfield  {author} {\bibinfo {author} {\bibfnamefont {X.}~\bibnamefont
  {Ren}}, \bibinfo {author} {\bibfnamefont {S.}~\bibnamefont {Frick}}, \bibinfo
  {author} {\bibfnamefont {A.}~\bibnamefont {McMillan}}, \bibinfo {author}
  {\bibfnamefont {S.}~\bibnamefont {Chen}}, \bibinfo {author} {\bibfnamefont
  {A.}~\bibnamefont {Halimi}}, \bibinfo {author} {\bibfnamefont {P.~W.}\
  \bibnamefont {Connolly}}, \bibinfo {author} {\bibfnamefont {S.~K.}\
  \bibnamefont {Joshi}}, \bibinfo {author} {\bibfnamefont {S.}~\bibnamefont
  {Mclaughlin}}, \bibinfo {author} {\bibfnamefont {J.~G.}\ \bibnamefont
  {Rarity}}, \bibinfo {author} {\bibfnamefont {J.~C.}\ \bibnamefont
  {Matthews}},  \emph {et~al.},\ }in\ \href@noop {} {\emph {\bibinfo
  {booktitle} {CLEO: Applications and Technology}}}\ (\bibinfo {organization}
  {Optical Society of America},\ \bibinfo {year} {2020})\ pp.\ \bibinfo {pages}
  {AM3K--6}\BibitemShut {NoStop}%
\bibitem [{\citenamefont {Morimoto}\ \emph {et~al.}(2021)\citenamefont
  {Morimoto}, \citenamefont {Wu}, \citenamefont {Ardelean},\ and\ \citenamefont
  {Charbon}}]{morimoto2021superluminal}%
  \BibitemOpen
  \bibfield  {author} {\bibinfo {author} {\bibfnamefont {K.}~\bibnamefont
  {Morimoto}}, \bibinfo {author} {\bibfnamefont {M.-L.}\ \bibnamefont {Wu}},
  \bibinfo {author} {\bibfnamefont {A.}~\bibnamefont {Ardelean}}, \ and\
  \bibinfo {author} {\bibfnamefont {E.}~\bibnamefont {Charbon}},\ }\href@noop
  {} {\bibfield  {journal} {\bibinfo  {journal} {Physical Review X}\ }\textbf
  {\bibinfo {volume} {11}},\ \bibinfo {pages} {011005} (\bibinfo {year}
  {2021})}\BibitemShut {NoStop}%
\bibitem [{\citenamefont {Valencia}\ \emph {et~al.}(2005)\citenamefont
  {Valencia}, \citenamefont {Scarcelli}, \citenamefont {D'Angelo},\ and\
  \citenamefont {Shih}}]{valencia_twophoton_2005}%
  \BibitemOpen
  \bibfield  {author} {\bibinfo {author} {\bibfnamefont {A.}~\bibnamefont
  {Valencia}}, \bibinfo {author} {\bibfnamefont {G.}~\bibnamefont {Scarcelli}},
  \bibinfo {author} {\bibfnamefont {M.}~\bibnamefont {D'Angelo}}, \ and\
  \bibinfo {author} {\bibfnamefont {Y.}~\bibnamefont {Shih}},\ }\href {\doibase
  10.1103/PhysRevLett.94.063601} {\bibfield  {journal} {\bibinfo  {journal}
  {Phys. Rev. Lett.}\ }\textbf {\bibinfo {volume} {94}},\ \bibinfo {pages}
  {063601} (\bibinfo {year} {2005})}\BibitemShut {NoStop}%
\bibitem [{\citenamefont {Gasparini}\ \emph {et~al.}(2018)\citenamefont
  {Gasparini}, \citenamefont {Zarghami}, \citenamefont {Xu}, \citenamefont
  {Parmesan}, \citenamefont {Garcia}, \citenamefont {Unternährer},
  \citenamefont {Bessire}, \citenamefont {Stefanov}, \citenamefont {Stoppa},\
  and\ \citenamefont {Perenzoni}}]{ISSCC2018}%
  \BibitemOpen
  \bibfield  {author} {\bibinfo {author} {\bibfnamefont {L.}~\bibnamefont
  {Gasparini}}, \bibinfo {author} {\bibfnamefont {M.}~\bibnamefont {Zarghami}},
  \bibinfo {author} {\bibfnamefont {H.}~\bibnamefont {Xu}}, \bibinfo {author}
  {\bibfnamefont {L.}~\bibnamefont {Parmesan}}, \bibinfo {author}
  {\bibfnamefont {M.~M.}\ \bibnamefont {Garcia}}, \bibinfo {author}
  {\bibfnamefont {M.}~\bibnamefont {Unternährer}}, \bibinfo {author}
  {\bibfnamefont {B.}~\bibnamefont {Bessire}}, \bibinfo {author} {\bibfnamefont
  {A.}~\bibnamefont {Stefanov}}, \bibinfo {author} {\bibfnamefont
  {D.}~\bibnamefont {Stoppa}}, \ and\ \bibinfo {author} {\bibfnamefont
  {M.}~\bibnamefont {Perenzoni}},\ }in\ \href {\doibase
  10.1109/ISSCC.2018.8310202} {\emph {\bibinfo {booktitle} {2018 IEEE
  International Solid - State Circuits Conference - (ISSCC)}}}\ (\bibinfo
  {year} {2018})\ pp.\ \bibinfo {pages} {98--100}\BibitemShut {NoStop}%
\bibitem [{\citenamefont {Kim}\ \emph {et~al.}(2021)\citenamefont {Kim},
  \citenamefont {Park}, \citenamefont {Chun}, \citenamefont {Choi},\ and\
  \citenamefont {Kim}}]{kim2021}%
  \BibitemOpen
  \bibfield  {author} {\bibinfo {author} {\bibfnamefont {B.}~\bibnamefont
  {Kim}}, \bibinfo {author} {\bibfnamefont {S.}~\bibnamefont {Park}}, \bibinfo
  {author} {\bibfnamefont {J.-H.}\ \bibnamefont {Chun}}, \bibinfo {author}
  {\bibfnamefont {J.}~\bibnamefont {Choi}}, \ and\ \bibinfo {author}
  {\bibfnamefont {S.-J.}\ \bibnamefont {Kim}},\ }in\ \href@noop {} {\emph
  {\bibinfo {booktitle} {2021 IEEE International Solid-State Circuits
  Conference (ISSCC)}}},\ Vol.~\bibinfo {volume} {64}\ (\bibinfo {organization}
  {IEEE},\ \bibinfo {year} {2021})\ pp.\ \bibinfo {pages}
  {108--110}\BibitemShut {NoStop}%
\end{thebibliography}%
$\,$\\
\noindent \textbf{Acknowledgements.} DF is supported by the Royal Academy of Engineering under the Chairs in Emerging Technologies scheme and acknowledges financial support from the UK Engineering and Physical Sciences Research Council (grants EP/T00097X/1 and EP/R030081/1). This project has received funding from the European union's Horizon 2020 research and innovation programme under the Marie Skłodowska-Curie grant agreement No. 754354. H.D. acknowledges support from the European Union's Horizon 2020 research and innovation programme under the Marie Sklodowska-Curie grant agreement No. 840958.  \\
\\
\noindent \textbf{Authors contributions.} D.F. and E.C. conceived the research. H.D., A.L. and J.Z. designed the experimental setup. J.Z. and A.L. performed the experiment. J.Z. and A.L. analysed the data. A.U. E.C. developed ~\textit{SwissSPAD2} and H.D. developed the coincidence counting algorithm. E.C. supervised the project. All authors contributed to the manuscript.\\
\\
\noindent \textbf{Data availability.} The experimental data and codes that support the findings presented here are available from the corresponding authors upon reasonable request.

\clearpage
\newpage
\onecolumngrid
\section*{Supplementary Information} 
\subsection{Details on data processing}

\begin{figure}[h]
\includegraphics[width=1 \columnwidth]{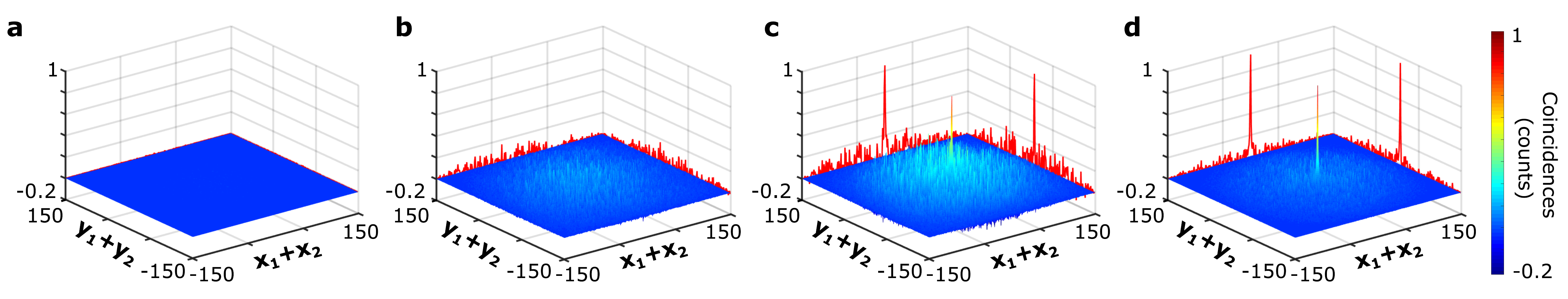}
\caption{\label{Figure5} \textbf{Full spatially-averaged correlation images.} \textbf{(a), (b), (c) and (d)} The full $300 \times 300$ spatially-averaged correlation images at 4 gate positions corresponding to those in Fig.~\ref{Figure2}. The red curves on the walls are the projections of the data across the correlation peak.} 
\end{figure}

The intensity images shown in this work are pixel-wise summed by the acquired number N frames. The value of hot pixels is set to be 0 for correlation calculation and then interpolated with the neighboring pixels to show a better intensity image. The spatially-averaged correlation image shown in the context is cropped  from the full $300 \times 300$ data for better visualization purpose. The results shown in Fig.~\ref{Figure5} are the full spatially correlation data corresponding to the 4 selected gate positions in Fig.~\ref{Figure2}. In Fig.~\ref{Figure5}c and d, the correlation peaks are well visible, while the background fluctuation in Fig.~\ref{Figure5}d is smaller as no classical light is collected at that gate position~\cite{defienne2019quantum}.

\begin{figure}[h]
\includegraphics[width=0.9 \textwidth]{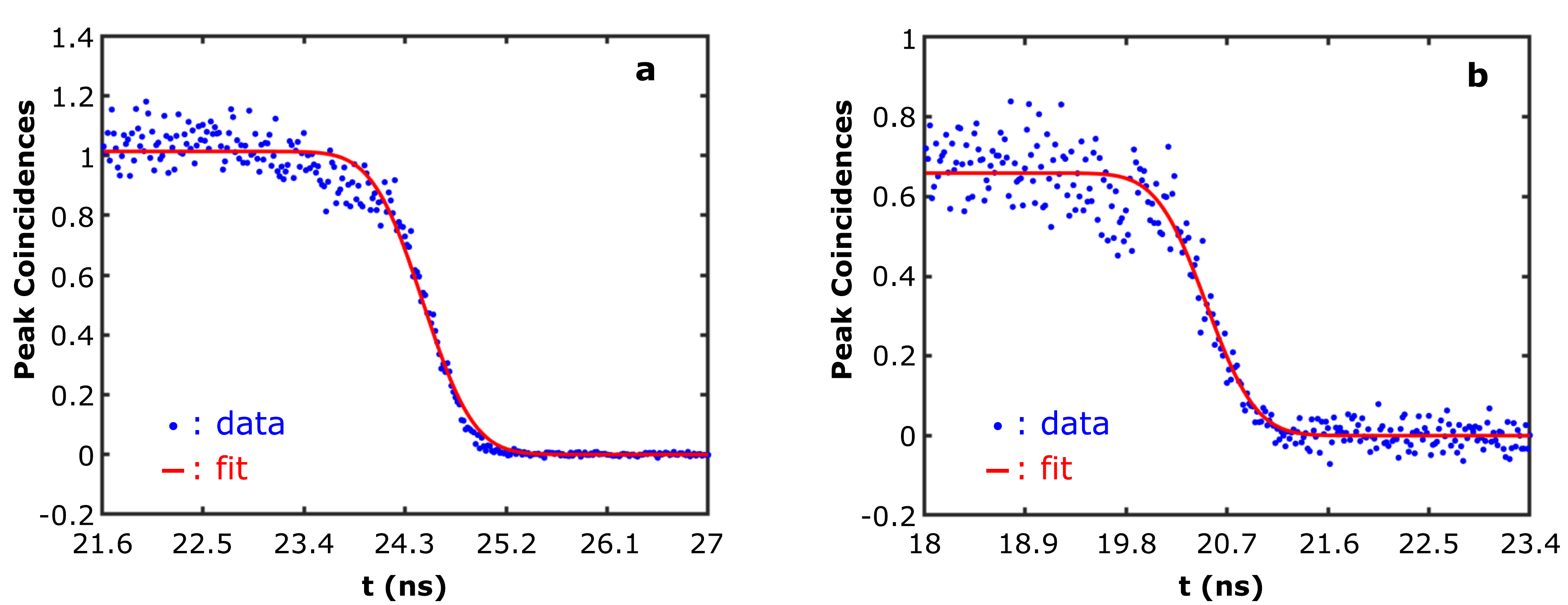} 
\caption{\label{Figure6} \textbf{Correlation peak data and fitted curves at falling edge.} \textbf{(a)} The correlation peak data from $21.6$ ns to $27$ ns corresponding to the synchronous result in Fig.~\ref{Figure2} and its fitted falling edge. \textbf{(b)} The correlation peak data from $18$ ns to $23.4$ ns corresponding to the asynchronous result in Fig.~\ref{Figure4} and its fitted falling edge. }
\end{figure}

The time arrival information of the quantum illuminated object can be located by finding the falling edge of the correlation peak profile. In order to locate the middle point at the falling edge, we fitted the fluctuated correlation peak data by the the error function $erf(.)$~\cite{chan2019long}. From the fitted curves shown in Fig.~\ref{Figure6}a and b, we obtained the falling time of $0.882$ ns for the synchronous case and $0.864$ ns for the asynchronous case (from $90\%$ to $10\%$). The tiny variation (1 gate position) also proves the reliability of the correlation peak profile used for object ranging. As the falling edge is not perfectly sharp, we recovered the intensity images of the classical object and quantum object by subtracting the intensity image at the gate position $45$ gates before the middle point from the one $45$ gates after. This allows us to achieve a subtracted intensity image with better contrast by avoiding using the images at the falling edge. However, the falling edge profile can be improved by optimizing the electrical gate shape of the camera and decreasing the width of the laser pulse.

\subsection{Correlation peak SNR analysis}

\begin{figure}[h]
\includegraphics[width=0.9 \textwidth]{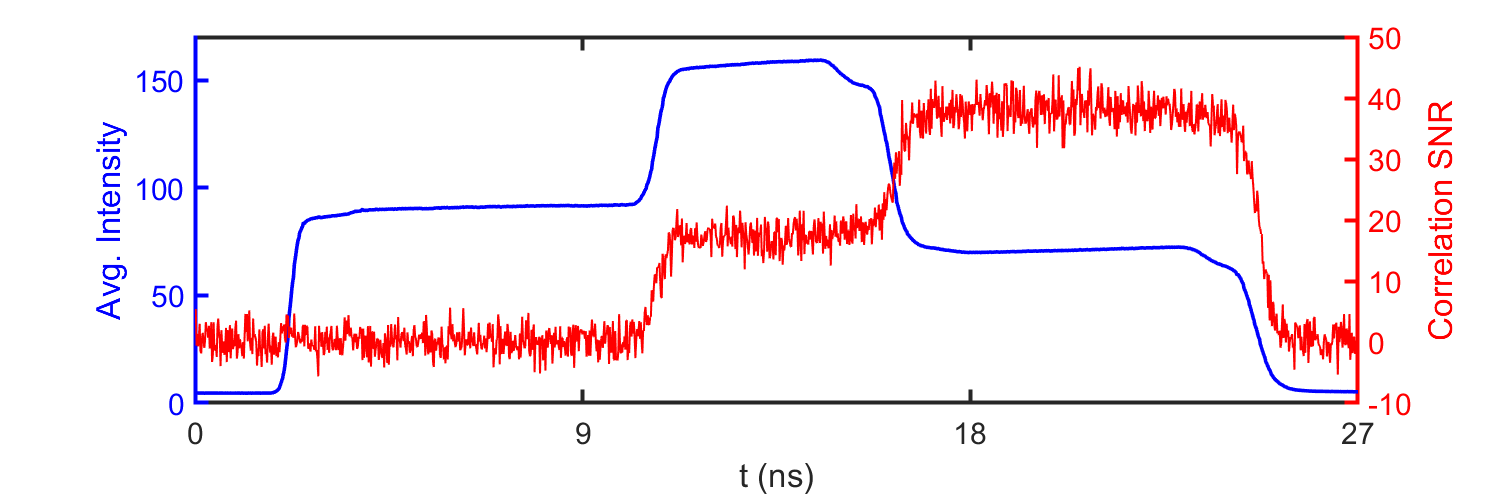} 
\caption{\label{Figure7} \textbf{Correlation peak SNR profile.} The SNR data of the synchronous result corresponding to Fig.~\ref{Figure2}. The average intensity profile is plotted as reference. }
\end{figure}

To analyse the visibility of the correlation peak in the spatially-averaged correlation image, we calculate the single-to-noise ratio (SNR) defined as the correlation peak value divided by the standard deviation of the background noise surrounding it. Fig.~\ref{Figure7} shows the SNR of the corresponding result in Fig.~\ref{Figure2} and the average intensity plotted as reference. We observe that the SNR increases above $1$ when quantum light arrives at the camera $11$ ns (classical light already present). In addition, the SNR is further improved when the classical light disappears at $17$ ns because the background noise decreases.

\begin{figure}[h]
\includegraphics[width=1 \textwidth]{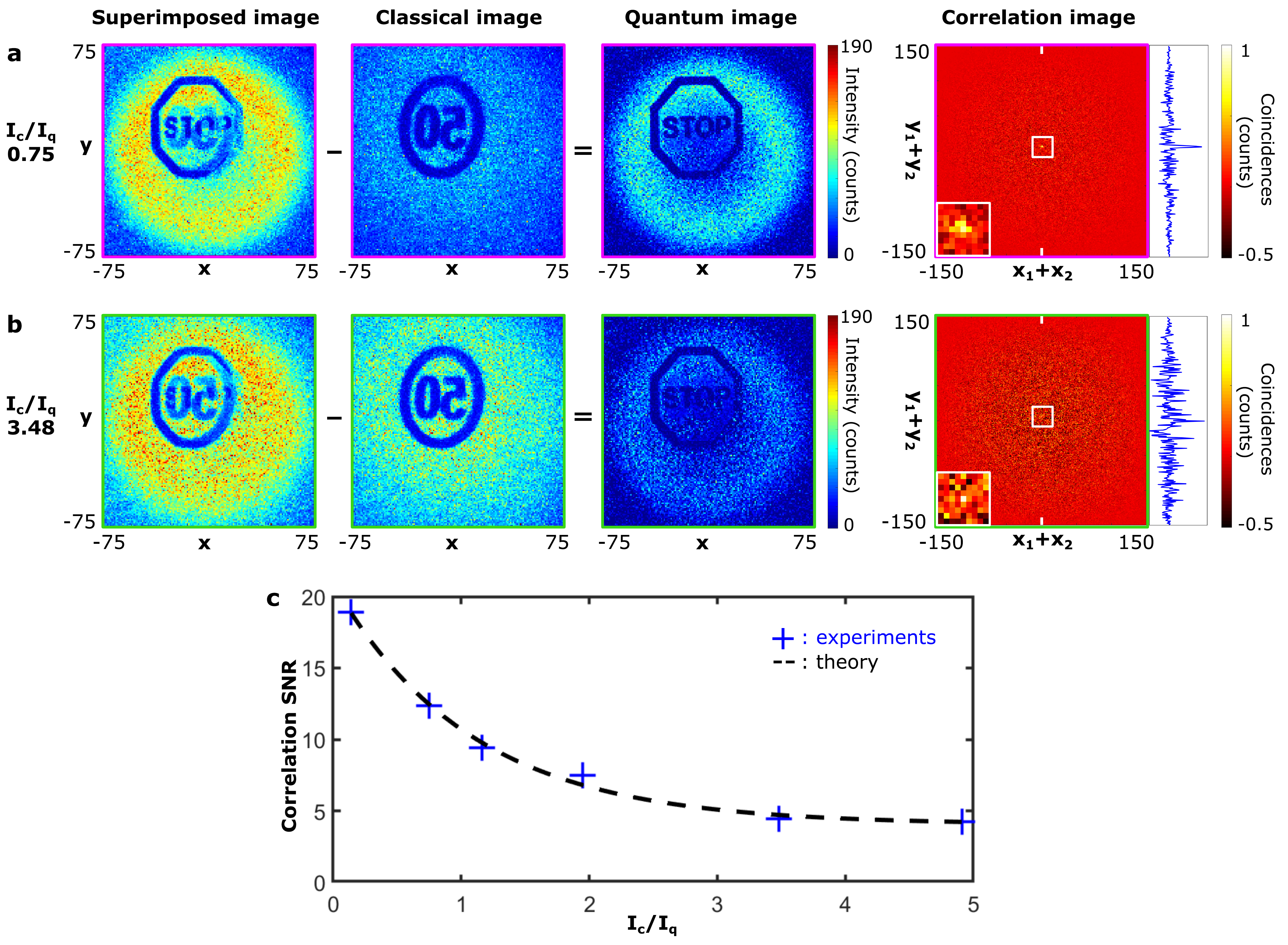} 
\caption{\label{Figure8} \textbf{SNR analysis with various classical over quantum light intensity ratios.} \textbf{(a)} The intensity images and spatially-averaged correlation images with $I_c / I_q = 0.75$. \textbf{(b)} The intensity images and spatially-averaged correlation images with $I_c / I_q = 3.48$. The insets in the correlation image are the central $9 \times 9$ data and the curves are the vertical profiles across the center. \textbf{(c)} More measured correlation SNR values with different $I_c / I_q$ ratios and the fitted curve with a theoretical model.}
\end{figure}

In order to further evaluate the visibility of the correlation peak in the spatially-average correlation image, we measure the correlation peak SNR for various classical light over quantum light intensity ratios ($I_c / I_q$) by keeping the pump laser power constant and tuning the classical laser power. The experiments are performed with the same camera configurations as in Fig.~\ref{Figure4}. As shown in Fig.~\ref{Figure8}a and b, the superimposed images are acquired at $13.5$ ns gate position where the gate window captures both classical light and quantum light. It is difficult to distinguish the traffic stop sign when $I_c/I_q$ is high ($3.48$) in the superimposed image, While it is better resolved after subtracting the classical light acquired at gate position of $24.66$ ns. From the spatially-averaged correlation image we can see that the background is more noisy when classical light intensity is higher as shown in the profiles across the center. Fig.~\ref{Figure8}c shows the measured correlation peak SNR values with various $I_c/I_q$ and the fitting curve with the theoretical model described in~\cite{defienne2019quantum}. All the results are based on the measurements with fixed $N=3000$ 8-bit frames. 

\begin{figure*}
\includegraphics[width=1 \textwidth]{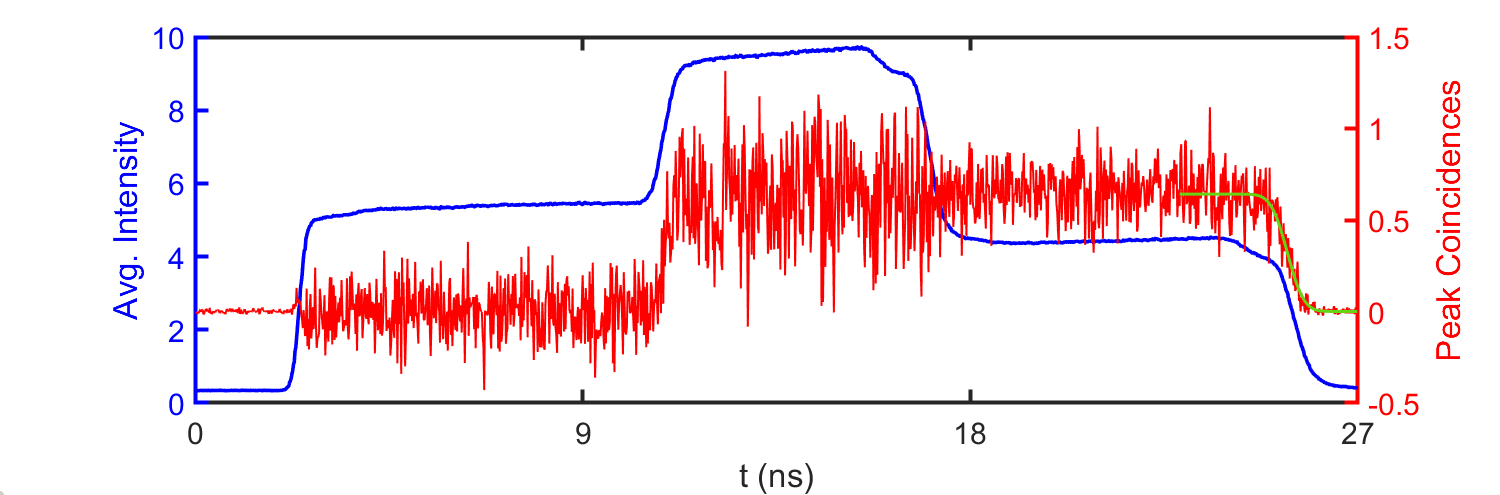} 
\caption{\label{Figure9} \textbf{Results with reduced number of frames per gate.} The average intensity and the correlation peak over 1500 gate positions are measured corresponding to the scenario in Fig.~\ref{Figure2}. At each gate position, $N=300$ frames are acquired with an acquisition time of $0.81$ seconds. The fitted curve of the correlation peak falling edge is shown with green line.}
\end{figure*}

\begin{figure*}
\includegraphics[width=1 \textwidth]{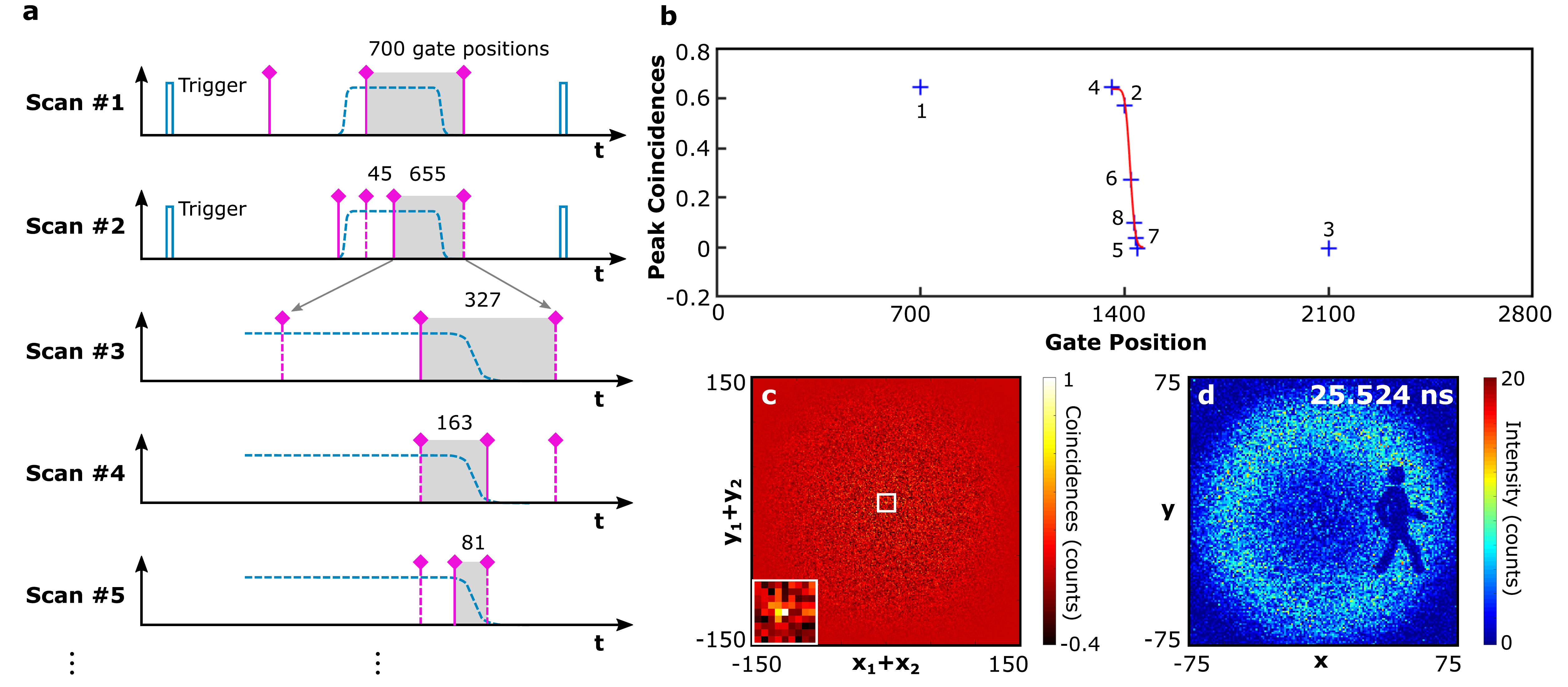} 
\caption{\label{Figure10} \textbf{Correlation-driven scanning approach and the result.} \textbf{(a)} The correlation-driven scanning principle. \textbf{(b)} Only $8$ scanning gate positions (300 frames acquired per gate) are required to locate the range of the object, which costs about $6.56$ seconds. \textbf{(c)} The spatially-averaged correlation image at scanning position 2 in \textbf{(b)}. The inset is the central $9 \times 9$ data. \textbf{(d)} shows the subtracted intensity image and the arrival time of the reflected quantum light from the object, $25.524$ ns.}
\end{figure*}

\subsection{Improving the quantum LiDAR acquisition speed}

\subsubsection{Reducing the number of frames} 
In Figure~\ref{Figure2}, $5000$ frames are acquired in $13.5$ s at each gate position to measure a spatially-averaged correlation image and identify the peak with a SNR on the order of $30$. As shown in Figure~\ref{Figure9}, the peak is still well visible (SNR on the order of $6$) if the number of acquired frame per gate is reduced to $300$, which strongly lower the acquisition time down to $0.81$ s per gate. 

\subsubsection{Using a correlation-driven algorithm} 
To further improve the quantum LiDAR speed, we developed a coincidence-driven algorithm inspired by the binary search process used in successive-approximation register (SAR) analog-to-digital converter. To cover the range of the object, the scanning time range should be larger than the laser pulse period. Here, we use $2800$ gate positions corresponding to $50.4$ ns. As shown in Fig.~\ref{Figure10}a, we initially scan $3$ gate positions with $700$ gates ($12.6$ ns) interval dividing the scanning range to $4$ parts equally. As the width of the gate window is $15.066$ ns, we can make sure that at least one of the initially scanned gates can capture the reflected quantum light pulse, thus a corresponding higher correlation peak will be obtained. Since the falling edge is not perfectly sharp, a following scanning process is implemented to check if the gate position with higher correlation peak is from the falling edge. To avoid such false locating, we scanned the gate position $45$ gates before and $45$ gates after the target position from the last scanning. The target range defined for the next scanning is between the last gate position with higher correlation peak and the scanned gate position just after it. When the target range is narrow enough the fitting method can be applied to achieve the falling edge according to the discrete scanned points. An example is depicted in Fig.~\ref{Figure10}b, where only $6.56$ seconds is consumed with $8$ scanning points to locate the range of the object. Fig.~\ref{Figure10}c shows the projected coincidences at scanning point 2, in which the coincidence peak is obvious. Fig.~\ref{Figure10}d is the subtracted intensity image and the measured relative range is $25.524$ ns. Note that the initial offset for measurements in Fig.~\ref{Figure2} and Fig.~\ref{Figure10} are different resulting different relative time ranges. 

By reducing the number of acquired frames to $300$ and using a correlation-driven algorithm, we retrieve the quantum-illuminated object and its depth in $7$ seconds. In the current implementation, it is not necessary to consider the gating profile variation over different pixels since the object is based on a 2 dimensional mask, which simplifies the processing and makes the proposed algorithm effective. However, the correlation-driven algorithm can be also extended to quantum LiDAR applications with 3 dimensional objects by applying in-pixel successive approximation. The similar approach has been implemented for conventional TCSPC-based LiDAR to reduce the output bandwidth~\cite{kim2021}.

\subsection{Analysis of different scenarios}

\begin{figure}[h]
\includegraphics[width=1 \textwidth]{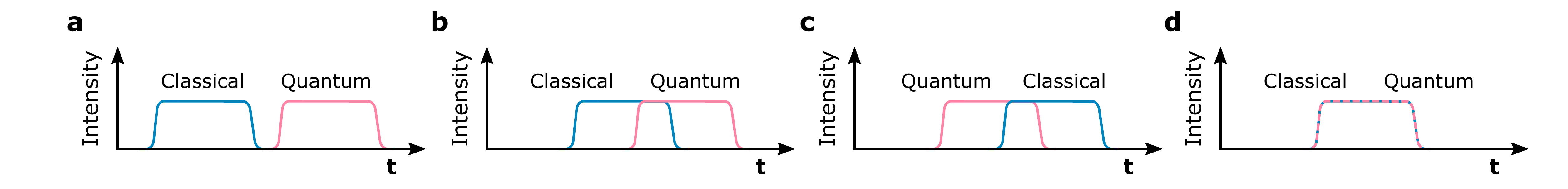} 
\caption{\label{Figure11} \textbf{Different synchronous scenarios.} \textbf{(a)} The response profiles of classical light and quantum light are separated. \textbf{(b)} The response profiles of classical light and quantum light are overlap, while the classical light arrives into the camera first similar to the scenario in Fig.~\ref{Figure2}. \textbf{(c)} The response profiles of classical light and quantum light are overlap, while the quantum light arrives into the camera first. \textbf{(d)} The two profiles are entirely overlap or the distance is smaller than the depth resolution of the camera, which can be only distinguished with spatially-resolved correlation image as shown in Fig.~\ref{Figure3}.}
\end{figure}

According to the different arrival time of the reflected photons from the classical light and quantum light, different scenarios are shown in Fig.~\ref{Figure11}. For the first three scenarios shown in Fig.~\ref{Figure11}a, b and c, the target quantum object can be located by the falling edge of the correlation peak profile and subtracted from the intensity image after the falling edge no matter it is background light (in a and b) or classical light (c). However, if the distance between the two classical and quantum objects are smaller than the depth resolution of the camera, the spatially-resolved correlation image has to be performed for distillation, which takes more time. 

\end{document}